\documentclass[letterpaper]{aa}  
\usepackage{graphicx}
\usepackage{txfonts}
\begin{document}
\title{Quasar induced galaxy formation:\\ a new paradigm ?}

   \author{D. Elbaz \inst{1}
          \and  K. Jahnke \inst{2}
         \and  E. Pantin \inst{1}
         \and  D. Le Borgne \inst{3,4}
         \and  G. Letawe \inst{5}
          }

   \institute{Laboratoire AIM, CEA/DSM-CNRS-Universit\'e Paris Diderot, IRFU/Service d'Astrophysique, B\^at.709, CEA-Saclay, 91191 Gif-sur-Yvette C\'edex, France\\
              \email{delbaz@cea.fr}
         \and Max-Planck-Institut f\"ur Astronomie, K\"onigstuhl 17, D-69117 Heidelberg, Germany
\and UPMC Univ Paris 06, UMR7095, Institut d'Astrophysique de Paris, F-75014, Paris, France
\and CNRS, UMR7095, Institut d'Astrophysique de Paris, F-75014, Paris, France
\and Institut d'Astrophysique et G\'eophysique, Universit\'e de Li\`ege,  All\'ee du 6 Ao\^ut, 17 Sart Tilman (Bat. B5C), B-4000 Li\`ege, Belgium  
             }

   \date{Received Juy 9, 2009; accepted September 18, 2009}

 
\abstract
{}
{We discuss observational evidence that quasars play a key role in the formation of galaxies starting from the detailed study of the quasar HE0450$-$2958 and extending the discussion to a series of converging evidence that radio jets may trigger galaxy formation.}
{We use mid infrared imaging with VISIR at the ESO-VLT, model the mid to far infrared energy distribution of the system and the stellar population of the companion galaxy using optical VLT-FORS spectroscopy. The results are combined with optical, CO, radio continuum imaging from ancillary data.}
{The direct detection with VISIR of the 7 kpc distant companion galaxy of HE0450$-$2958 
allows us to spatially separate the sites of quasar and star formation activity in this composite system made of two ultra-luminous infrared galaxies (ULIRGs), where the quasar makes the bulk of the mid infrared light and the companion galaxy powered by star formation dominates in the far infrared. No host galaxy has yet been detected for this quasar, but the companion galaxy stellar mass would bring HE0450$-$2958 in the local M$_{\rm BH}$--M$_{\star}^{\rm bulge}$ relation if it were to merge with the QSO. This is bound to happen because of their close distance (7 kpc) and small relative velocity ($\sim$60--200 km s$^{-1}$). We conclude that we may be witnessing the building of the M$_{\rm BH}$--M$_{\star}^{\rm bulge}$ relation, or at least of a major event in that process. The star formation rate ($\sim$340 M$_{\sun}$yr$^{-1}$), age (40--200 Myr) and stellar mass ($\sim$[5--6]$\times$10$^{10}$ M$_{\sun}$) are consistent with jet-induced formation of the companion galaxy. We suggest that HE0450$-$2958 may be fueled in fresh material by cold gas accretion from intergalactic filaments. We map the projected galaxy density surrounding the QSO as a potential tracer of intergalactic filaments and discuss a putative detection.
Comparison to other systems suggests that inside-out formation of quasar host galaxies and jet-induced galaxy formation may be a common process. Two tests are proposed for this new paradigm: (1) the detection of offset molecular gas or dust emission with respect to the position of distant QSOs, (2) the delayed formation of host galaxies as a result of QSO activity, hence the two step building of the M$_{\rm BH}$/M$_{\star}^{\rm bulge}$ ratio.
}
{}
\keywords{Galaxies: active --
                Galaxies: formation --
                Galaxies: jets --
                quasars: individual: HE0450-2958
               }

\maketitle
\section{Introduction}
The correlation that exists between the mass of the central supermassive black holes (SMBH) of local galaxies and their bulge luminosity (Kormendy \& Richstone 1995, Magorrian et al. 1998), stellar mass (McLure \& Dunlop 2001, 2002, Kormendy \& Gebhardt 2001, Marconi \& Hunt 2003, Ferrarese et al. 2006) or velocity dispersion (Gebhardt et al. 2000, Ferrarese \& Merritt 2000), suggests the existence of a physical mechanism connecting the activity of a galaxy nucleus and the bulding of its host galaxy stellar bulge. The search for this mechanism is one of the major goals of observational and theoretical astronomy since it may bring a key element in our global understanding of galaxy formation. AGN or QSO activity is often invoked to explain the origin of red-dead galaxies in the local Universe and the downsizing effect (massive galaxies formed their stars first, see e.g. Cowie et al. 1996 and discussion in Sect.~\ref{SEC:roleQSO}), by quenching star formation in galaxies, e.g. after the merging of two gas rich massive spiral galaxies  (Di Matteo et al. 2005, Croton et al. 2006, Schawinski et al. 2006). Mergers may both trigger star formation and provoke the loss of angular momentum required to feed a central SMBH (Sanders et al. 1988, Comerford et al. 2009) that would then eject the remaining gas fuel of the galaxy. However, even if QSOs ultimately stop star formation in galaxies, they may as well act as the prime triggering mechanism in their past star formation history, through positive feedback due to radio jets and turbulent pressure (see e.g. Silk \& Norman 2009, Silk 2005, Begelman \& Cioffi 1989, Feain et al. 2007). As such they would provide an alternative mechanism to galaxy mergers to explain the physical triggering of starbursts, the short timescale associated to the formation of stars in galaxy bulges (see e.g. Thomas et al. 2005) and provide a natural explanation for the enhanced Mg/Fe ratio observed in ellipticals (Pipino, Silk \& Matteucci 2009). Even if active nuclei accelerated star formation at early stages, they may still later on quench it.

The aim of the present study is to give credit to AGN/QSO positive feedback and to suggest that the formation of whole galaxies may have been provoked by quasars. We report observations in the mid infrared of the QSO HE0450$-$2958, which may represent an early phase in a scenario of "quasar induced galaxy formation". HE0450$-$2958 is a nearby luminous ($M_V$$=$$-25.8$) radio quiet quasar located at a redshift of $z$=0.2863 (Canalizo \& Stockton 2001) and presently the only known quasar for which no sign of a host galaxy has been found (Magain et al. 2005). Originally, a mass of $M_{\rm BH}$$\sim$8$\times$10$^8$ M$_{\odot}$ was derived by Magain et al. (2005) for its central SMBH, assuming that its accretion rate was about half the Eddington limit for a radiation efficiency of $\eta$=0.1. Adopting the local M$_{\rm BH}$--M$_{\star}^{\rm bulge}$ relation of McLure \& Dunlop (2002) to derive the stellar mass of its host galaxy and then converting it into a V-band absolute magnitude (-23.5$\leq$ $M_V$ $\leq$-23.0, depending on the age of the stellar population), Magain et al. (2005) found that such massive host galaxy should have been 6 to 16 times brighter than the upper limit set by the V-band HST-ACS image of the quasar. This upper limit of  -21.2$\leq M_V \leq$-20.5 was obtained after deconvolution of the V-band HST-ACS image of the quasar using the MCS deconvolution technique (Magain, Courbin \& Sohy 1998), with the benefit of the presence of a nearby star only 2\arcsec\,away from the quasar. Only diffuse photoionized gas -- including a very closeby cloud that they qualified as "the blob" -- and a 7 kpc (1.5\arcsec) distant "companion galaxy", were identified in the vicinity of the quasar. 

This is not the first time that the existence of so-called "naked" quasars has been claimed but earlier findings (Bahcall, Kirhakos \& Schneider 1994, 1995, Bahcall et al. 1997) were disproved after smoothing the HST high resolution images in which low surface brightness host galaxies could not be detected (McLeod \& Rieke 1995). However, in the case of HE0450$-$2958, even using a  two-dimensional fit of the HST image of HE0450$-$2958, the host galaxy remained unseen (see Kim et al. 2007, who derived an upper limit consistent with the initial one of Magain et al. 2005). HE0450$-$2958 has since been the center of a debate and various scenarios have been proposed to explain the absence of detection of a host galaxy for this system. Here we recall those scenarios and propose an alternative one, in which the companion galaxy participates to the building of the stellar mass of the future host galaxy of HE0450$-$2958.

We use recently obtained near-infrared data from the Very Large Telescope Imager and Spectrometer in the Infrared (VISIR) of the European Southern Observatory (ESO) Very Large Telescope (VLT) to demonstrate that HE0450$-$2958 is a composite system made of a pair of ultra-luminous infrared galaxies (ULIRGs), the QSO and its companion galaxy where most, if not all, star formation is taking place. 

In Sect.~\ref{SEC:composite}, we present the VISIR observations at 11.3\,$\mu$m. We show that both the QSO and its companion galaxy are directly detected. We model the spectral energy distribution (SED) in the infrared of both systems, from which a star formation rate (SFR) is derived for the companion galaxy. 

In Sect.~\ref{SEC:companion}, we compute the stellar population age and mass of the companion galaxy from the model fit of its optical spectrum from the VLT-FORS at ESO. We suggest that this galaxy may be newly formed and present a scenario to explain its formation, as a result of the radio jets of the neighboring QSO.

In Sect.~\ref{SEC:scenario}, we discuss four scenarios to explain the reason why no host galaxy have yet been found associated with the QSO HE0450$-$2958. We propose that the QSO and its companion galaxy will merge together, hence the companion galaxy may be considered as 	a candidate future host galaxy for HE0450$-$2958.
Even if a yet undetected host galaxy was present at the location of the QSO, the companion galaxy would anyway participate to the mass building of the final host galaxy.

In Sect.~\ref{SEC:origin}, we propose a mechanism to explain the quasar activity of HE0450$-$2958, even in the absence of a host galaxy, based on the accretion of cold gas filaments as inferred to explain the bulk of star formation in galaxies by Dekel et al. (2009). 

In Sect.~\ref{SEC:roleQSO}, we discuss the role of quasars in galaxy formation in general. We list a series of seven converging evidence that quasars participate to galaxy formation and discuss some of them in detail. We extract two observational tests of the "quasar induced galaxy formation" scenario proposed here that will come out from future observations of QSOs.

In Sect.~\ref{SEC:conclusion}, we summarize the results of the paper and the main elements suggesting that radio jets may represent the missing link connecting active nuclei and galaxy formation.

Throughout we will use Vega zero-points and a cosmology of
$h=H_0/(100~\mathrm{km s^{-1} Mpc^{-1}})=0.7$, $\Omega_M=0.3$, and
$\Omega_\Lambda=0.7$, corresponding to linear scales of 4.3 kpc/\arcsec at $z=0.286$.

   \begin{figure*}[ht!]
   \centering
   \includegraphics*[width=18cm]{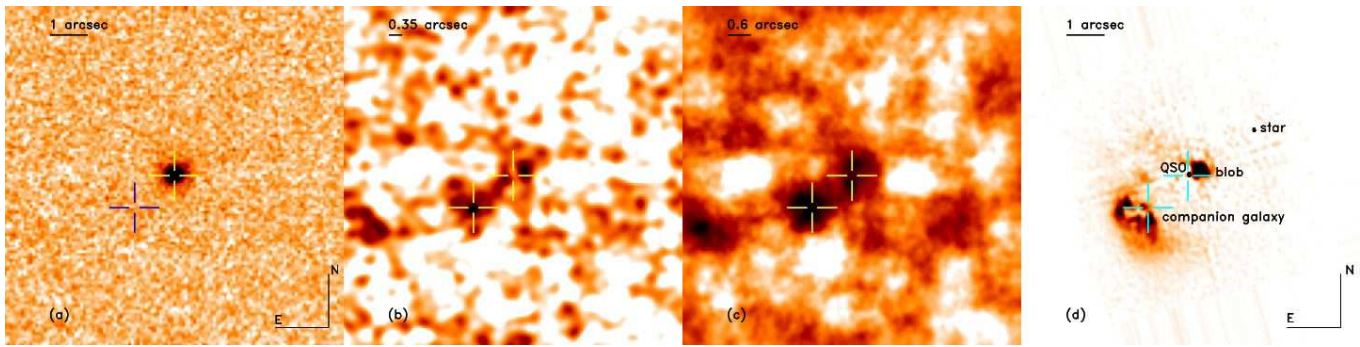}
      \caption{\textbf{\textit{(a)}}VISIR image of HE0450$-$2958 ($z$=0.2863) in the PAH2 (11.3\,$\mu$m) filter, combining two exposures of 1623 seconds. The central source is consistent with a point source with the FWHM of VISIR at this wavelength of 0.35\arcsec. \textbf{\textit{(b)}} VISIR image after PSF subtraction of the central source followed by a convolution with the VISIR PSF. A second source is clearly visible on the S-E of the QSO (1.5\arcsec\,away), at the position of the companion galaxy of HE0450$-$2958, as well as a residual emission to the N-W of the QSO, coincident with the blob of gas seen in the HST-ACS image. \textbf{\textit{(c)}} The PSF-subtracted VISIR image after applying a mean-filtering with an aperture of 0.6\arcsec\,of radius. In both the middle and right images, the brightest source after PSF-subtraction of the QSO is located to the S-E of the center of the image, at the position of the companion galaxy. \textbf{\textit{(d)}} HST-ACS image after deconvolution of the QSO by Magain et al. (2005). The blue crosses show the central position of the QSO used for the PSF-subtraction in the VISIR image and MCS-deconvolution in the HST-ACS image and the position of the S-E source detected at the 5.1-$\sigma$ level in the VISIR image. The point source to the N-W of the QSO in the HST-ACS image is a star that is not detected at 11.3\,$\mu$m by VISIR.}
         \label{FIG:3visir}
   \end{figure*}

\section{HE0450$-$2958: a composite system made of a pair of ULIRGs}
\label{SEC:composite}
A bright source was detected by IRAS at the approximate location of HE0450$-$2958 within the position error bar of 5\arcmin. At the luminosity distance of the QSO ($z$=0.2863), its flux densities of 69.3 ($\pm$ 19.4), 188.8 ($\pm$ 6.9), 650  ($\pm$ 52) and 764.6  ($\pm$ 12.2) mJy in the IRAS 12, 25, 60 and 100\,$\mu$m passbands respectively (IRAS Faint Source Catalog, Moshir et al. 1990) translate into the total infrared luminosity of an ultra-luminous infrared galaxy (ULIRG, L$_{\rm IR}$=L(8--1000\,$\mu$m)$\geq$10$^{12}$ L$_{\sun}$, de Grijp et al. 1987, Low et al. 1988). Using the H$\alpha$/H$\beta$ ratio as an indicator of dust obscuration for the QSO and a neighboring galaxy, that they labeled as the "companion galaxy" (see Fig.~\ref{FIG:3visir}d), Magain et al. (2005) claimed that the source of the IRAS emission was not the QSO but instead this galaxy, hence that the absence of a host galaxy could not be justified by dust obscuration.

To test this hypothesis, we imaged HE0450$-$2658 with VISIR, the Very Large Telescope Imager and Spectrometer in the Infrared (VISIR) at the ESO-VLT to identify at a sub-arcsec scale the position of the mid infrared emitter and see whether it was associated to the QSO or companion galaxy. The companion galaxy is located at nearly the same redshift ($z$=0.2865, Canalizo \& Stockton 2001) than the QSO and at a projected separation of 1.5\arcsec (7 kpc) in the S-E direction, hence well beyond the point spread function (PSF) full width half-maximum (FWHM) of VISIR of 0.35\arcsec (see Fig.~\ref{FIG:3visir}). A first analysis of this dataset was presented in Jahnke et al. (2009) discussing the near and mid infrared emission of the QSO. Here we show that both the QSO and its companion galaxy are detected with VISIR.

\subsection{VISIR observations and data reduction}
\label{SEC:visir}
The VISIR data on HE0450$-$2658 (P.I. K.Jahnke, ESO program 276.B-5011(A)) cover a field of view of 9\arcsec\,(120 pixels). The pixel scale is 0.075\arcsec\,and the exposure time is 1623 seconds for each subset. The combined image has a point source sensitivity of 3 mJy (5$\sigma$). To suppress the large mid infrared background, we used the standard chopping-nodding observing scheme with a chopper throw of 8\arcsec\,that maintains the source always in the detector field of view. Observations were performed with the PAH2 filter centered at 11.3\,$\mu$m. At this wavelength, the point spread function (PSF) has a full width half-maximum (FWHM) of 0.35\arcsec, i.e. 860 times sharper than IRAS. The data reduction was done following the technique discussed in Pantin et al. (2007) and the flux calibration was done using two reference stars (HD 29085, 4.45 Jy and HD 41047, 7.21 Jy). 
In a first analysis (Jahnke et al. 2009), only a bright source was detected associated with the QSO. Here we present a refined analysis combining two (instead of one) VISIR datasets to obtain an improved image analysis allows us to detect two sources, the second one being associated with the companion galaxy of HE0450$-$2658. We test the robustness of our results using both 1623 seconds images either in separated or in combined form.
\begin{table}
\caption{Optical position and redshift of the QSO and companion galaxy in the HE0450$-$2958 system.}
\label{TAB:data}
\begin{tabular}{cccc}
  \hline
  Source ÊÊ&ÊRedshift$^{\mathrm{(a)}}$ Ê&ÊRa$^{\mathrm{(b)}}$  Ê&ÊDec$^{\mathrm{(b)}}$ \\
  \hline
QSO & 0.2863 & 04h52m30.105s& -29$\degr$53$\arcmin$35.57$\arcsec$ \\
Companion & 0.2865 & 04h52m30.201s& -29$\degr$53$\arcmin$36.58$\arcsec$\\
  \hline
\end{tabular}
\begin{list}{}{}
\item[$^{\mathrm{(a)}}$]Reference: Canalizo \& Stockton (2001).
\item[$^{\mathrm{(b)}}$]HST--ACS positions (J2000).
\end{list}
\end{table}
 
\begin{table*}
\caption{Optical to far infrared photometry of the HE0450$-$2958 system (QSO and companion galaxy).}
\label{TAB:photometry}
\centering                                    
\begin{tabular}{c||c cc||ccccc}
  \hline
    Source ÊÊ   &ÊACS$^{\mathrm{(a)}}$        & WFPC2$^{\mathrm{(a)}}$   & NICMOS$^{\mathrm{(a)}}$  & VISIR$^{\mathrm{(b)}}$  (mJy)           &Ê                        & IRAS$^{\mathrm{(b)}}$ (mJy)               &                     & \\
               ÊÊ   & F606W    &  F702W   & F160W      &11.3\,$\mu$m &12\,$\mu$m      & 25\,$\mu$m     & 60\,$\mu$m & 100\,$\mu$m \\
  \hline
QSO            & 15.46       &   15.23     & 13.6         & 62.5$\pm$0.9  & 69.3$\pm$19.4& 188.8$\pm$6.9& 650$\pm$52& 765$\pm$122 \\
Companion & 18.69       &   17.88     & 15.77       &   4.5$\pm$0.9  &   -                     &   -                     &    -               &   -                   \\
  \hline
\end{tabular}
\begin{list}{}{}
\item[$^{\mathrm{(a)}}$]Photometric measurements in the optical (HST--ACS 606 nm, HST--WFPC2 702 nm) and near infrared (HST--NICMOS 1.6\,$\mu$m, from Jahnke et al. 2009) are given in Vega magnitudes. AB magnitudes can be computed using AB - Vega values of 0.07 (ACS-F606W), 0.23 (WFPC2-F702W), 1.31 (NICMOS-F160W). 
\item[$^{\mathrm{(b)}}$]Mid to far infrared flux densities in the VISIR and IRAS passbands are given in mJy.
\end{list}
\end{table*}
\subsection{Results of the VISIR observations: detection of the companion galaxy of HE0450$-$2958}
\label{SEC:resvisir}
The raw image shows a very bright source visible at the very center of the VISIR image (Fig.~\ref{FIG:3visir}a), centered on the optical position of the QSO (see Table~\ref{TAB:data}). The QSO emission is detected with a flux density of 62.5 mJy, hence more than two orders of magnitudes above the rms level of 0.6 mJy for point source detection. Its flux is so large that it must be masked or PSF-subtracted in order to see other objects detected in the VISIR image at the 5-$\sigma$ level. In Figs.~\ref{FIG:3visir}a,b we present the results of two image processing techniques applied on the co-added image (for a total integration time of 3246 seconds on source) after PSF-subtraction of the central bright source associated with the QSO. 

In Fig.~\ref{FIG:3visir}b, we apply an optimal filtering by convolving the residual image (after PSF-subtraction of the central brightest source) with the VISIR PSF in order to optimize the contrast for point sources in the image. Note that the PSF removal at the position of the brighest source, i.e. the QSO, helps to improve the contrast of the structures for the eye but is not critical in the quantitative analysis of the companion galaxy. We did not perform a Standard Image deconvolution to the VISIR image because it is an inverse problem which is not currently applicable to VISIR images due to slight PSF variations. Instead, the convolution with the PSF (so-called optimal filtering) is not really affected by PSF instabilities (direct process, low-pass filtering).

The brightest source in the resulting image is located in the S-E direction with respect to the QSO at the location of the companion galaxy (see Fig.~\ref{FIG:3visir}d, see also Fig.~\ref{FIG:contours}a). Another source appears in the residual image, to the N-W of the central QSO position, which coincides with the position of a gas "blob" seen in the optical image. The optical spectrum of this gas blob exhibits no evidence for the presence of stars and its origin remains unknown (Magain et al. 2005). If confirmed, the VISIR source associated with it would imply that it also contains dust, hence that it may have been ejected by the QSO. We also note the marginal detection of emission acting as a bridge between the QSO and the companion galaxy which can be seen in blue contours in Fig.~\ref{FIG:contours}a. If confirmed from deeper imaging, this detecion may be associated with material expelled by the QSO, hence participating to the jet.

\begin{figure}
   \centering
   \includegraphics*[width=8cm]{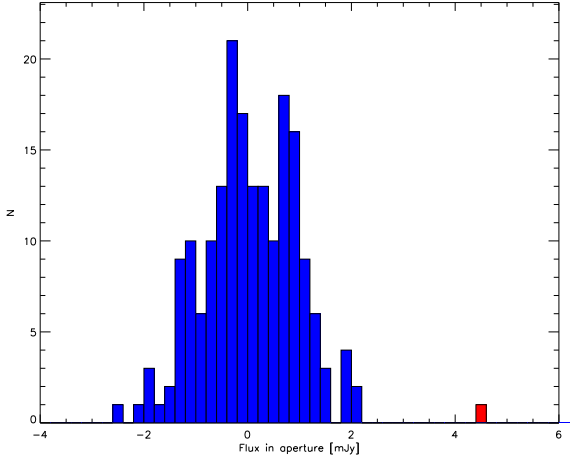}
      \caption{Histogram of aperture measurements in the VISIR image (Fig.~\ref{FIG:3visir}) within a 0.6\arcsec\,radius. Measured values were multiplied by an aperture correction factor for point sources with the VISIR PSF of 1.17. Only positions distant by 0.6\arcsec\,in both the Ra and Dec directions are presented (196 positions or pixels in Fig.~\ref{FIG:3visir}c). The only position located away from the nearly gaussian distribution of all values is colored in red. It coincides with the position of the companion galaxy of HE0450$-$2958, located 1.5\arcsec\,to the S-E of the QSO.
      }
         \label{FIG:histo}
   \end{figure}

In the second image processing technique (Fig.~\ref{FIG:3visir}c), we apply a mean filtering, i.e. each image pixel is replaced by the sum of the flux densities measured in an aperture of 0.6\arcsec\,radius. With the VISIR PSF, an aperture correction factor of 1.17 then needs to be applied to obtain the total flux density of point sources. The brightest source is again located at the optical position of the companion galaxy. In order to quantify the robustness of this detection, we plot in Fig.~\ref{FIG:histo} the histogram of all aperture measurements obtained in the image with a separation of  0.6\arcsec\,both in the Ra and Dec directions (196 positions) to avoid duplicated measurements. The distribution of the measurements follows a gaussian shape as expected for a purely white noise, with an rms of 0.89 mJy. The only aperture measurement away from the noise distribution corresponds to the position of the companion galaxy. The exact position of the brightest aperture measurement was found to be located only 0.06\arcsec\,away from the centroid of the companion galaxy in the HST-ACS image (see Table~\ref{TAB:data}). It is detected with a flux density of 4.5 mJy at the 5.1-$\sigma$ level at 11.3\,$\mu$m in the observed frame, hence 8.9\,$\mu$m in the rest-frame. There is marginal evidence that the source detected at the position of the companion galaxy is extended as can be seen in Fig.~\ref{FIG:3visir}b,c. 

The VISIR and HST-ACS images (after MCS deconvolution, from Magain et al. 2005) of HE0450$-$2958 are compared in Fig.~\ref{FIG:3visir}d at the same scale and with the same orientation showing that the two VISIR sources are coincident with the two optical sources, the QSO and its 7 kpc distant companion galaxy. In the VISIR images Fig.~\ref{FIG:3visir}b and c, the QSO was PSF subtracted. The second source is distant from the QSO by more than four times the FWHM of the PSF (0.35\arcsec, labeled with an horizontal line in Fig.~\ref{FIG:3visir}b), hence the source is not a residual after the PSF-subtraction of the QSO nor an artefact due to an anomalously elongated PSF, which is not typically found in VISIR observations anyway. 

We checked that the source associated with the companion galaxy of HE0450$-$2958 in the combined VISIR image was also detected in both individual 1623 seconds images before and after PSF-subtracting the QSO, hence the source cannot be due to noise fluctuations, since the noise would have to produce the same fake object in both images. We produced a histogram of aperture measurements from both images, as in Fig.~\ref{FIG:histo}, both after and before PSF-subtraction of the QSO (avoiding the central part of the image contaminated by the emission from the QSO) and found the same result as in Fig.~\ref{FIG:histo} with a signal-to-noise ratio of $\sim$3.5 in the individual images. 

\subsection{The energy balance of the HE0450 system: separation of star formation and accretion}
\label{SEC:balance}
   \begin{figure}
   \centering
   \includegraphics*[width=9.3cm]{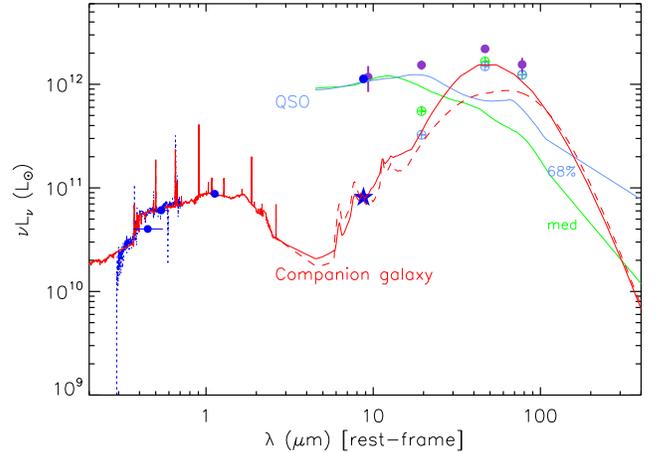}
      \caption{Full spectrum of the companion galaxy of HE0450$-$2958 from 0.3 to 400\,$\mu$m in the rest-frame. The optical/near-infrared part of the figure compares the optical model fit (red line) to the VLT-FORS spectrum (dashed blue line) using PEGASE.2 (discussed in Sect.~\ref{SEC:companion} and Fig.~\ref{FIG:FORS}. Broadband measurements are summarized in Table~\ref{TAB:photometry}. The filled blue dots in the optical range come from the HST measurements in the F606W, F702W, F160W passbands of ACS, WFPC2, NICMOS for the companion galaxy. The filled purple circles show the IRAS measurements in the four passbands centered at 12, 25, 60 \& 100\,$\mu$m and filled blue symbols, the VLT--VISIR detections at 11.3\,$\mu$m (observed; 8.9\,$\mu$m rest-frame) associated to the QSO (filled blue circle) and companion galaxy (filled blue star). The plain green and blue lines represent the median and 68 percentile reddest SEDs from the Atlas of 47 quasar SEDs of Elvis et al. (1994), normalized to the VISIR measurement. Empty circles with crosses mark the expected emission of the companion in the IRAS passbands after subtraction of the two template quasar SEDs. The plain and dashed red lines in the mid to far infrared are the Chary \& Elbaz (2001) SEDs adjusted to the VISIR and IRAS emission (plain line) and to the VISIR emission of the companion galaxy alone (dashed line). The green and blue dashed lines show the combined model SED of the QSO plus companion galaxy (fit to VISIR and IRAS residuals) going through the IRAS measurements which do not resolve both objects.
              }
         \label{FIG:combinedSED}
   \end{figure}
   \begin{figure*}[ht!]
   \centering
   \includegraphics[width=6cm]{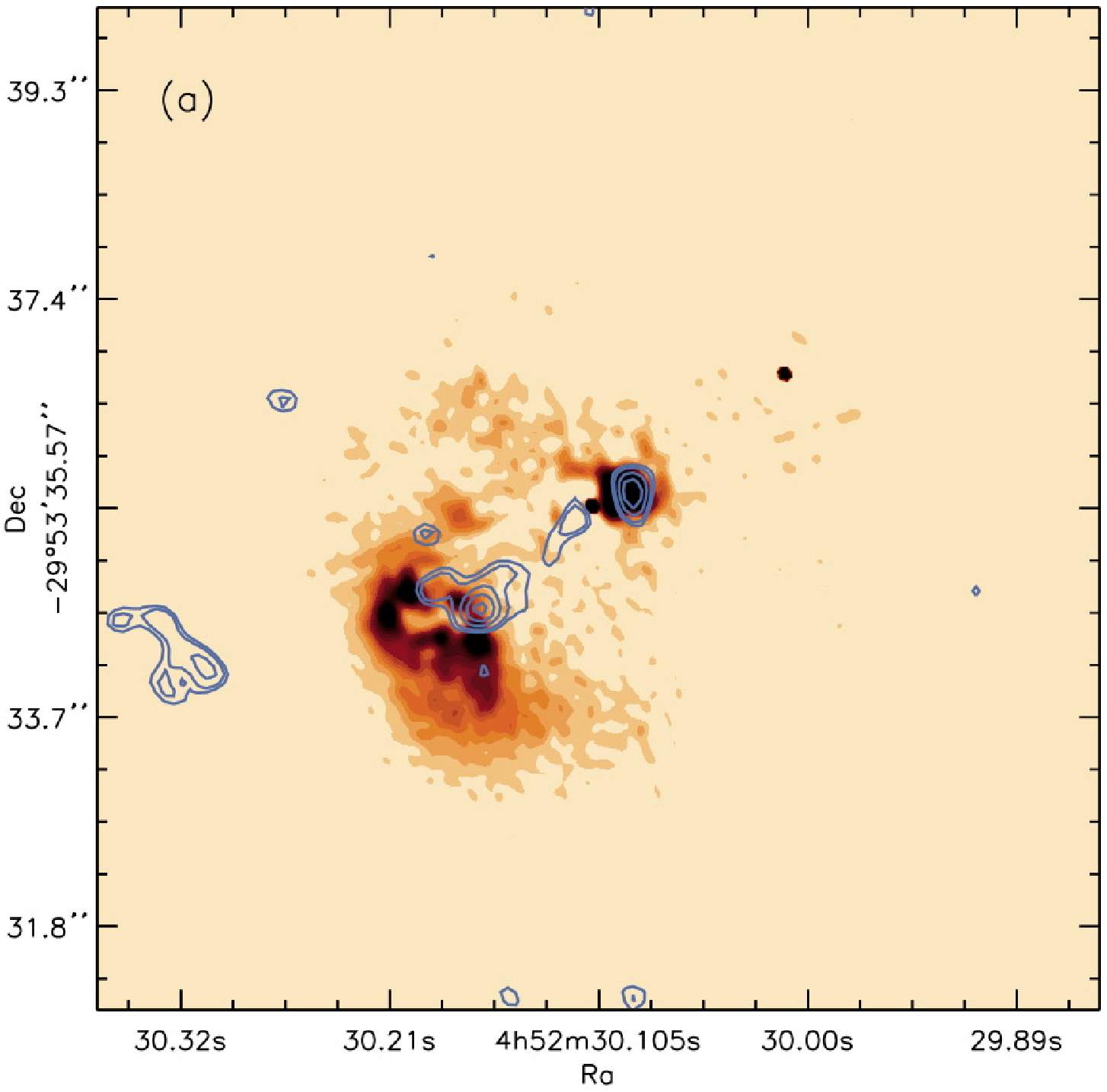} 
   \includegraphics[width=6cm]{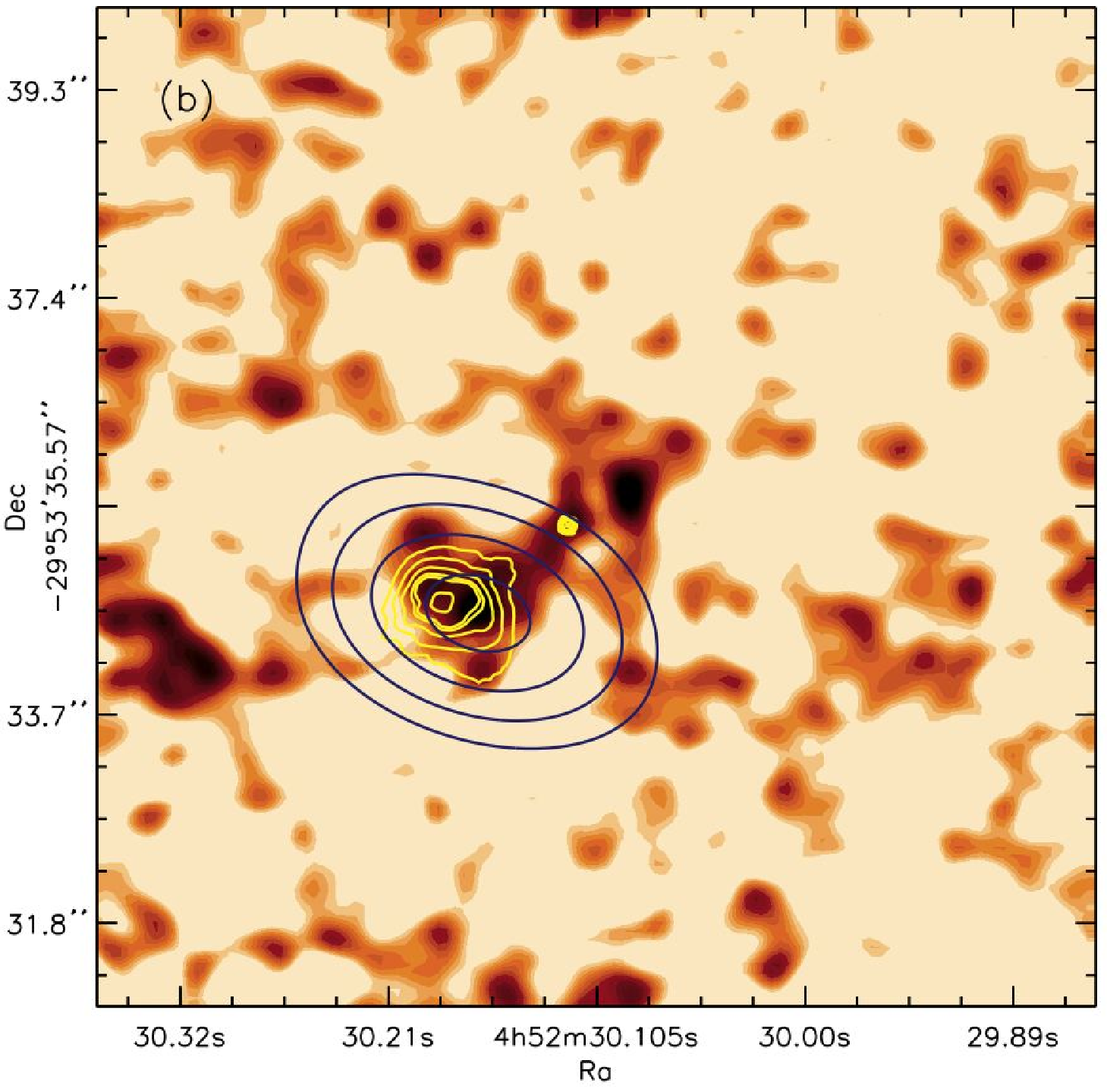} 
   \includegraphics[width=6cm]{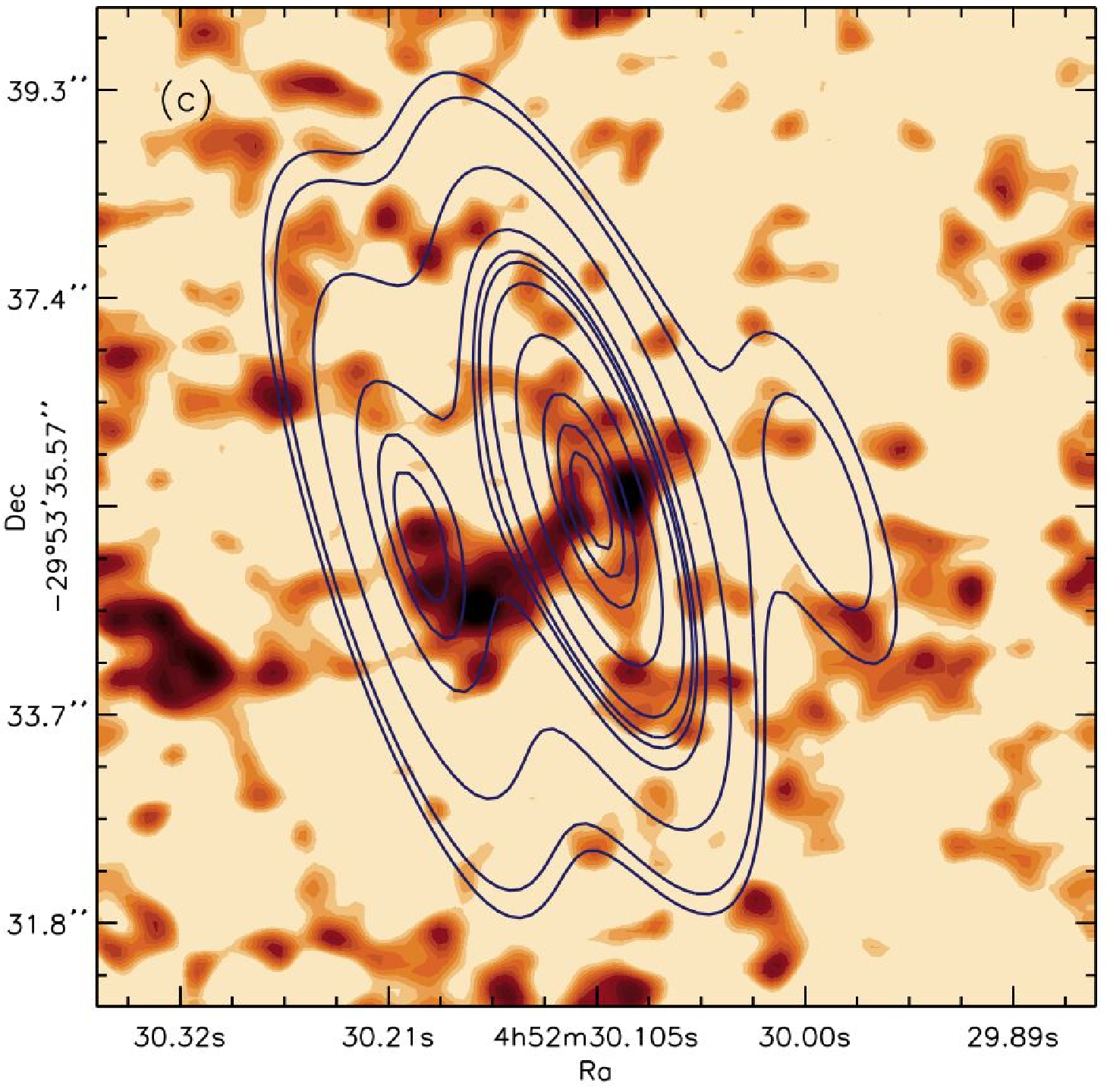} 
      \caption{\textbf{\textit{(a)}} VISIR 11.3\,$\mu$m contours (in blue, 8.9\,$\mu$m in the rest-frame) overlayed on the HST-ACS image obtained after PSF deconvolution of the QSO by Magain et al. (2005). The VISIR image used for the contours is the one of Fig.~\ref{FIG:3visir}b where the point source QSO was PSF-subtracted and an optimal filtering was applied (convolution with PSF). Note the extended structure from the QSO to the companion galaxy which may result from dust associated with the jet and the detection at the position of the blob to the N-W suggesting the presence of dust in it. This VISIR image is shown in \textbf{\textit{(b)}} together with contours from NICMOS-F160W at 1.6\,$\mu$m (in yellow, 1.2\,$\mu$m in the rest-frame, Jahnke et al. 2009) and ATCA CO 1--0 (dark blue lines, from Papadopoulos et al. 2008a) contours. In \textbf{\textit{(c)}}, the ATCA 6208 MHz radio continuum (dark blue lines, from Feain et al. 2007) emission associated to the radio jets is overlayed on the VISIR image.}
         \label{FIG:contours}
   \end{figure*}
The VISIR image provides the first direct evidence that the IRAS "source" is in fact made of two well separated objects, the QSO and its companion galaxy. The infrared SED of the "system" (the QSO and companion galaxy are both contained in the single IRAS measurement) peaks around $\sim$50\,$\mu$m (rest-frame) which indicates that the bulk of the far infrared light arises from star formation. This is a natural result from the fact that centrally concentrated infrared emission associated with active nuclei exhibit very high dust temperatures of hundreds of kelvins which consequently peak in the mid infrared. Indeed the Spitzer QUEST sample of quasars shows that quasar spectra drop beyond $\sim$20\,$\mu$m that Netzer et al. (2007) interpret as the signature of a minimum temperature of $\sim$200 K.

For comparison, the median SED of a sample of 47 quasars (Elvis et al. 1994, see their Fig.11) normalized to the VISIR 11.3\,$\mu$m luminosity of HE0450$-$2958 exhibits a rapid fall at higher wavelengths (green line in Fig.~\ref{FIG:combinedSED}), which cannot be reconciled with the peak IRAS emission at $\sim$50\,$\mu$m (rest-frame). To illustrate the variety of quasar infrared SEDs, we also show the shape of the 68 percentile reddest SED (plain blue line) from this atlas of quasar SEDs which again can only marginally account for a larger fraction of the far infrared emission, mostly at 20\,$\mu$m. At a redshift of $z$=0.2863, both SEDs give a comparable total infrared luminosity for the QSO of L$_{\rm IR}$=L[8--1000\,$\mu$m] $\sim$ 1.7 and 2.2$\times$10$^{12}$ L$_{\sun}$ for the median and 68\,\% reddest SED respectively.

It is possible to separate the contribution of both sources in the mid infrared thanks to VISIR, but this remains to be done for the far infrared part where the SED peaks. The VISIR 11.3\,$\mu$m (observed) measurements for the QSO (62.5 mJy) and companion galaxy (4.5 mJy) of $\nu$L$\nu$(8.9\,$\mu$m)= 97.2 and 9.3$\times$10$^{10}$ L$_{\sun}$ respectively at their redshifts of $z$=0.2863 and 0.2865 are marked with a filled blue circle and star in Fig.~\ref{FIG:combinedSED}. The sum of the two sources detected by VISIR at 11.3\,$\mu$m (67 mJy), agrees very well with the IRAS measurement at 12\,$\mu$m of 69.3$\pm$ 19.4 mJy, the slight excess in the IRAS passband (well within the IRAS error bar) may be explained by the slightly larger central wavelength of the passband in the rising part of the SED. We estimate the contribution of star formation to the IRAS luminosities by subtracting the SED of Elvis et al. (1994) normalized to the VISIR 11.3\,$\mu$m emission of the QSO (open circles with crosses).  

We assume that the VISIR emission located at the position of the companion galaxy is uniquely due to star formation since no evidence has yet been found for the presence of an active nucleus in it (see Sect.~\ref{SEC:companion}). The fit of the VISIR 8.9\,$\mu$m (rest-frame) and IRAS far infrared luminosities (open circles with crosses in Fig.~\ref{FIG:combinedSED}) is obtained using a template SED (red plain line) from the library of local template SEDs in the infrared by Chary \& Elbaz (2001). We note that local galaxies with an 8.9\,$\mu$m luminosity like the one of the companion galaxy exhibit a median temperature of order 35-45 K, hence their far infrared SED peaks around 70-80\,$\mu$m instead of 50-60\,$\mu$m here. A warmer dust temperature of about 50-60 K, like the one of Arp 220, is required to explain the bulk of the far infrared emission for the companion galaxy. This is consistent with the optical and near-infrared morphology of the galaxy, with very opaque centrally concentrated regions producing the bulk of the near-infrared emission and nearly invisible in the optical. In Fig.~\ref{FIG:combinedSED}, we combine the infrared SED of the companion galaxy with the model spectrum that we use to fit its VLT-FORS optical spectrum normalized to its NICMOS $H$-band luminosity (the optical fit is discussed in Sect.~\ref{SEC:companion}). The total infrared luminosity of the companion galaxy is L$_{\rm IR}$$\sim$2$\times$10$^{12}$ L$_{\sun}$, for the best-fit SED which corresponds to a SFR$\sim$340 M$_{\sun}$yr$^{-1}$ assuming the conversion factor of 1.72$\times$10$^{-10}$ [M$_{\sun}$yr$^{-1}$] L$_{\sun}^{-1}$ of Kennicutt (1998) for a Salpeter Initial Mass Function.

Hence HE0450$-$2958 appears to be a composite system made of a pair of ULIRGs, where both mechanisms, star formation and QSO activity, are spatially separated by 7 kpc in two distinct sites. The question of the origin of the starburst in the companion galaxy is addressed in the next Section.

\subsection{Other sources of evidence that the companion of HE0450$-$2958 is a ULIRG: near infrared and CO imaging}
\label{SEC:ulirg}
Until now only indirect evidence had been found that the companion galaxy of HE0450$-$2958 was a ULIRG. This was initially claimed by Magain et al. (2005) on the basis of its Balmer decrement, but the attenuation could only be inferred from the part of the galaxy that shines in the optical suggesting that the companion galaxy suffered from dust extinction, but not necessarily that it was a ULIRG. 

The optical morphology of the companion galaxy of HE0450$-$2958, classified as a ring galaxy by Canalizzo \& Stockton (2001), is very peculiar with a few bright knots and what looks like a hole in the center (see Fig.~\ref{FIG:3visir}d). However, Jahnke et al. (2009) showed that the NICMOS-F160W 1.6\,$\mu$m image of the companion galaxy peaks at the center where no optical emission is seen (Fig.~\ref{FIG:contours}b), suggesting that the bulk of the dust obscured stellar light was originating from a region missed by optical spectroscopy. The centrally peaked emission in the near IR implies a very high level of extinction in the visible which suggests that a large fraction of the star formation activity taking place in this galaxy is indeed heavily dust obscured as in local ULIRGs, such as Arp220. 

A previous indication that the companion galaxy of HE0450$-$2958 was itself a ULIRG was provided by the CO map of the system (Papadopoulos et al. 2008a). The molecular gas traced by the CO molecule appears to avoid the QSO but to peak at the location of the companion galaxy (dark blue contours in Fig.~\ref{FIG:contours}b) close to the peak mid-infrared emission measured with VISIR. This was interpreted by Papadopoulos et al. (2008) as evidence that the companion galaxy was the locus of a strong star formation event usually associated with large molecular gas concentrations, while no evidence for such reservoir was found associated with the QSO. The mass of molecular gas (M(H$_2$)$\simeq$2.3$\times$10$^{10}$ M$_{\odot}$ assuming a conversion factor, X$_{\rm CO}$ of ULIRGs or M(H$_2$)$\simeq$1.25$\times$10$^{10}$ M$_{\odot}$ for an optically thin emission, Papadopoulos et al. 2008a) is equivalent to that of the super-antennae galaxy (a ULIRG of L(IR)=1.1$\times$10$^{12}$ L$_{\odot}$) and ten times larger than in the antennae galaxy (a LIRG of L(IR)=1.1$\times$10$^{11}$ L$_{\odot}$, Sanders \& Mirabel 1996). In the Milky Way, the color excess, E($B$-$V$), which measures the degree of dust extinction is spatially well correlated with molecular hydrogen, as traced by the CO emission (see e.g. Bohlin, Savage \& Drake 1978). Hence the fact that the CO emission in this field is centered on the companion galaxy suggests both that there is a large amount of dust in this galaxy and that there is a large gas reservoir to fuel an intense star formation event.

\section{The companion galaxy of HE0450$-$2958: a newly formed galaxy ?}
\label{SEC:companion}
   \begin{figure}
   \centering
   \includegraphics[width=9cm]{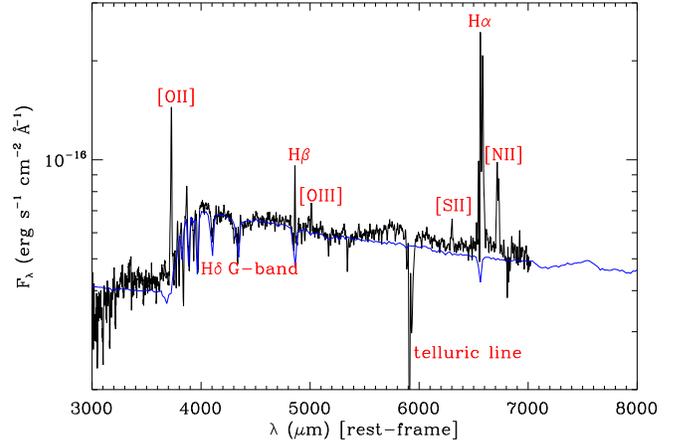}
      \caption{Fit of the ESO Very Large Telescope (VLT) spectrum obtained with FORS1 (Letawe et al. 2007), in black--bold line, using the model PEGASE.2 (Fioc \& Rocca-Volmerange 1997, 1999) in green--light line for a 200 Myr stellar population having formed with a star formation rate SFR$\sim$exp(-t/$\tau$), where $\tau$=100 Myr. Lower axis: rest-frame wavelengths (3000--7000\,\AA).
              }
         \label{FIG:FORS}
   \end{figure}
\subsection{Age and mass of the stellar population of the companion galaxy}
\label{SEC:stars}
The VLT--FORS optical spectrum of the companion galaxy (Letawe et al. 2008a) covers the 3000--7000\,\AA\,rest-frame wavelength range (see Fig.~\ref{FIG:FORS}). It contains strong emission lines of [OII], H$\beta$ and H$\alpha$ typical of young and massive stars as well as large equivalent width Balmer absorption lines typical of A stars with ages of a few 100 Myr and a moderate 4000\,\AA\,discontinuity. The origin of the [OIII], [NII], [SII] and [SIII] lines is discussed in Sect.~\ref{SEC:jet}. Those lines are affected by the nearby presence of the QSO and will provide important information for the discussion of the physical origin of the starburst in the companion galaxy. Here we fit the stellar continuum emission, hence unaffected by the presence of the QSO, to derive the age and stellar mass of the stellar population.

We fit the optical spectrum with the stellar population synthesis code PEGASE.2 (Fioc \& Rocca-Volmerange 1997, 1999). The best fit of the spectrum is obtained with a 200 Myr old stellar population having formed with a star formation rate SFR$\sim$exp[-t/$\tau$], where $\tau$=100 Myr (blue line in Fig.~\ref{FIG:FORS}). However this is not the only star formation history able to fit this optical spectrum. If instead of continuous star formation, we use a single burst, then the best fit is obtained for a single stellar population aged of $\sim$40 Myr assuming an extinction of $A_V\sim$1.7 (for a Calzetti law, Calzetti et al. 2000 with $R_V$=4.05). Older single stellar populations would produce a too large 4000\,\AA\, discontinuity, this is in particular the case for a 200 Myr single population (single burst). The dominant stellar population required to fit this spectrum is therefore rather young, i.e. a few 10 Myr, whether it was formed over a longer star formation history (SFR$\sim$exp[-t/100Myr]) or during a single burst. The presence of an older stellar population cannot be definitely excluded, as long as its contribution to the stellar continuum is largely diluted by the young stellar population. However no obvious indication for the presence of an old stellar population is found. In order to derive a stellar mass for the companion galaxy, we normalize the model optical spectra to the observed optical to near-infrared broadband photometry (see Table~\ref{TAB:photometry}). The stellar mass is M$_{\star}$=6$\times$10$^{10}$ M$_{\odot}$ for the continuous star formation history and 17\,\% lower, i.e. M$_{\star}$=5$\times$10$^{10}$ M$_{\odot}$, for the 40 Myr single stellar population. This mass is close to the one inferred by Merritt et al. (2006) and nearly as large as that of the Milky Way. Within the uncertainties in the derivation of the present-day SFR (SFR$\sim$340 M$_{\sun}$yr$^{-1}$, Sect.~\ref{SEC:balance}) and stellar mass of the galaxy, the present event of star formation is clearly the major event of star formation of this galaxy and may even be the first one.

\subsection{Origin of the starburst: merger-induced versus jet-induced star formation}
\label{SEC:jet}
We have seen in Sect.~\ref{SEC:composite} that HE0450$-$2958 is a composite system harboring both quasar activity and star formation, but in two separate locations, the quasar and the companion galaxy. In the local Universe, star formation rates as large as that of ULIRGs (SFR$>$170 M$_{\odot}$ yr$^{-1}$) are typically found during the merger of two nearly equal mass (ratio $>$ 1/4) massive spiral galaxies, i.e. major mergers (Ishida 2004, Sanders \& Mirabel 1996, Genzel, Lutz \& Tacconi 1998). In the absence of any massive host galaxy detected, Papadopoulos et al. (2008) suggested that HE0450$-$2958 might provide first evidence for another triggering mechanism for ULIRGs, i.e. the minor merger of a massive galaxy (the companion galaxy) with a dwarf elliptical galaxy. The latter would be the galaxy housing the quasar both below the optical detection limit and with faint or no CO emission. 

Here we wish to suggest an alternative possibility for the triggering of the ULIRG phase in the companion galaxy: jet-induced star formation. Evidence for such mechanism in other systems are presented in Sect.~\ref{SEC:jetinduced}. As discussed by Feain et al. (2007), the radio continuum map from the Australia Telescope National Facility (ATCA) shows a double-sided radio jet centered on the QSO (dark blue contours in Fig.~\ref{FIG:contours}c). The radio continuum emission cannot be due to emission by star formation from the companion galaxy itself since there is a slight offset of the radio jet with respect to the center of the companion galaxy. Moreover the presence of a symmetric jet in the opposite direction to that of the companion galaxy, with no obvious optical counterpart, and the connection between the ionized gas and radio emission also strongly favor the presence of radio jets in HE0450$-$2958 (Feain et al. 2007, Letawe et al. 2008a). 

Evidence for the influence of jet induced shocks can be found in the spectrum of the companion galaxy in the form of [OII], [OIII], [NII] and [SII] emission lines (Fig.~\ref{FIG:FORS}). Using a diagnostic diagram comparing [OIII]/H$\beta$ to [NII]/H$\alpha$, Letawe et al. (2008b, see their Fig.16c) found that while the gaseous regions surrounding the QSO HE0450$-$2958 exhibits high ionization levels due to photo-ionization from the QSO light, the companion galaxy occupies a very different locus in the diagram. Its large [NII]/H$\alpha$ implies other source of ionization than stellar radiation, however its low [OIII]/H$\beta$ ratio is not consistent with photo-ionization and suggests instead that these emission lines are shock-induced, implying that the intense star formation activity in the companion galaxy is triggered by the radio jet itself. Further evidence for such physical connection comes from the spatially resolved regions producing the [OIII] emission as measured with integral-field optical spectroscopy using VLT/VIMOS (Letawe et al. 2008a). Their Fig.13 presents a clear alignment between the radio continuum emission measured with ATCA by Feain et al. (2007) and the [OIII] emission. No such spatial association with the radio continuum is found for the regions responsible for the [OII] or H$\alpha$ emissions, which trace the excitation by stars. Interestingly, we see also marginal evidence for an alignment of the mid infrared map of the system from VISIR (blue contours in Fig.~\ref{FIG:contours}a). Lastly, a broad H$\alpha$ component was detected in the region lying between the QSO and companion galaxy with a speed of $\sim$900 km s$^{-1}$ moving away from the QSO and toward the companion galaxy, in a direction close to the plane of the sky (discussed in the Sect.6.6 of Letawe et al. 2008b). 

We conclude that the starburst event in the companion galaxy of HE0450$-$2958 is most probably triggered by the radio jets emitted by the neighboring QSO. We consider that although possible, it is highly improbable that such configuration happened by chance during a merger process. Several examples for jet-induced star formation are discussed and compared to this system in Sect.~\ref{SEC:jetinduced}. In the next Section, we discuss a formation scenario where the whole galaxy formation was induced by the radio jet of the QSO.

\subsection{Toward a scenario to explain the formation of the companion galaxy of HE0450$-$2958} 
\label{SEC:form_companion}
We propose here a scenario to explain the formation of the companion galaxy of HE0450$-$2958 through a process induced by the radio jets of the QSO. In this scenario, a pre-existing gas cloud is hit by the radio jet and star formation is induced in it leading to the formation of the companion galaxy as we see it now. An alternative scenario would be that the mass of the galaxy itself was expelled by the radio jet or that a galaxy randomly passed through the radio jet of the QSO. 

The relative velocity between the two galaxies is within the range 60--200 km s$^{-1}$ as derived from the emission lines in both optical spectra (Letawe et al. 2008b). Since a velocity of 1 km s$^{-1}$ is equivalent to 1 kpc Gyr$^{-1}$, during the lifetime of the radio jet, e.g. $\sim$0.1 Gyr, the QSO has moved by 6--20 kpc or 1.3--4.3\arcsec, at the redshift of the system. In the local Universe, the average galaxy density is $\phi^{\star}$=0.0213$h^3$= 0.0073 gal.Mpc$^{-3}$ (Croton et al. 2005), hence the average distance between two galaxies is $\sim$5 Mpc--comoving or 4 Mpc--proper at $z$=0.2863. The random collision of a QSO with another galaxy is therefore highly improbable, especially if it is required that it takes place when the radio jets are active, unless the QSO lies in a particularly dense region of the Universe that would favor such encounters. We mapped a region of $\sim$40 Mpc centered on HE0450$-$2958 to search for a particularly large galaxy density in its close neighborhood but we did not find any evidence for a particularly overdense region such as a massive galaxy cluster centered on the QSO (see Sect.~\ref{SEC:origin}). 

The companion galaxy is dominated by, if not uniquely made of, young stars aged of about 200 Myr, and possibly less (e.g. 40 Myr, see Sect.~\ref{SEC:stars}), and its star formation rate (SFR$\sim$340 M$_{\sun}$yr$^{-1}$, Sect.~\ref{SEC:balance}) is large enough to explain the formation of the total mass contained in those stars within the lifetime of the galaxy (M$_{\star}$$\sim$[5--6]$\times$10$^{10}$ M$_{\sun}$, Sect.~\ref{SEC:stars}). Compelling evidence for jet-induced star formation in the companion galaxy were presented in Sect.~\ref{SEC:jet}.  Hence it is possible that the companion galaxy is a newly formed galaxy which birth was triggered by the radio jet of the neighboring QSO. The dominant stellar population age is of the same order as typical radio jet lifetimes and assuming a radio jet speed of about 0.02c (see e.g. Scheuer 1995), it would only take 1 Myr for the radio jet to reach the 7 kpc distant companion galaxy. Hence timescales are consistent although the age of the radio jet is not known. This would make this system comparable to Minkowski's Object, a proto-typical case for jet-induced star formation (van Breugel et al. 1985). Minkowski's Object is a newly formed galaxy, with a stellar population age of only 7.5 Myr and a stellar mass of 1.9$\times$10$^7$ M$_{\odot}$ (Croft et al. 2006), aligned with the radio jet of NGC 541, an FR I galaxy (Fanaroff \& Riley 1974) located in the cluster Abell 194. The companion galaxy of HE0450$-$2958 might be a scaled-up version of Minkowski's Object, i.e. a massive galaxy which formation was jet-induced. 

We therefore propose that we are witnessing a process that we call "quasar-induced galaxy formation" through which the companion galaxy was triggered by the radio jets of the QSO HE0450$-$2958. We note the presence of three structures in the close vicinity of the quasar which may represent three successive steps in the process: (1) a blob of photoionized gas, adjacent to the quasar, deprived of stars (Magain et al. 2005), (2) an elongated gas $+$ stars (see with HST-NICMOS imaging) structure extending toward the N-E from the position of the quasar (Jahnke et al. 2009), (3) the companion galaxy itself dominated by a $\sim$40--200 Myr-old stellar population with distorted morphology. The presence of ionized gas moving from the quasar in the direction of the companion galaxy (Letawe et al. 2008b), along the radio jet, strengthens the idea that matter is being transported from the quasar to the companion galaxy. 

As noted by Feain et al. (2007), the radio jet of HE0450$-$2958 presents an offset in the northern direction with respect to the companion galaxy, which is seen in Fig.~\ref{FIG:contours}c. This offset implies that it is not the galaxy itself which is responsible for the whole radio emission on this side of the QSO. It also suggests that the companion galaxy was sweeped by the radio jet. Indeed the shape of the companion galaxy is extended in the South-North direction and its interstellar medium presents a gradient of ionization, ranging from lower ionization in the South to the highest levels of ionization in the Northern part of the galaxy, hence closer to the peak radio emission (Letawe et al. 2008a) This gradient is better explained if the ionization of the interstellar medium (ISM) in the companion is shock induced instead of being due to photoionization, in agreement with its emission line ratios. For those reasons, our interpretation is that the formation of the companion galaxy was shock induced by the radio jet of the QSO by sweeping a pre-existing cloud of ionized gas.

Evidence for the presence of large amounts of matter in the surrounding of radio quasars, in the form of massive gas clouds exhibiting strong emission lines with high excitation levels, has been gathered since the early 60's (Matthews \& Sandage 1963, Sandage \& Miller 1966, Wampler et al. 1975, Stockton 1976, Richstone \& Oke 1977). These extended emission lines regions (EELRs, Stockton \& MacKenty 1983) can reach masses of several 10$^{10}$ M$_{\odot}$ (Fu \& Stockton 2008, Stockton et al. 2007). Early guesses regarding the origin of EELRs were centered on tidal debris (Stockton \& MacKenty 1987) or cooling flows (Fabian et al. 1987), but ionized gas from a tidal encounter would dissipate on a timescale of less than a Myr (Crawford et al. 1988; see also discussion by Fu \& Stockton 2007 and the reviews by McCarthy 1993, Miley \& de Breuck 2008). Spectro-imaging and narrow-band filter imaging in the fields of radio quasars demonstrated that EELRS were made of clouds of gas, whose strong emission lines were excited by the hard radiation of the quasar. Fu \& Stockton (2008) found that low metallicity quasar host galaxies happened to be surrounded by EELRs with similar abundance ratios, which they interpreted as evidence that the EELRs had been expelled from the host galaxy itself. The double-lobed morphologies of extragalactic radio sources shows that the relativistic jets not only couple strongly with the ISM of the host galaxies, but are capable of projecting their power on the scales of galaxy haloes and clusters of galaxies (up to $\sim$1 Mpc, Tadhunter 2007). 

Several EELRs were found in the surrounding of HE0450$-$2958 itself following the axis of the radio jets. They have been detected up to 30 kpc away from the QSO (Letawe et al. 2008a). The N-W blob of gas, photoionized by the radiation of HE0450$-$2958, is itself located in the direction of the radio jet opposite to the one pointing at the companion galaxy. Hence we believe that the companion galaxy was formed after one of these gas clouds was sweeped by the radio jet of the QSO. The thermodynamic state of EELRs is uncertain since the densities one can infer from their optical spectra are of the order of n$_{\rm e}$=100--300 cm$^{-3}$ and T$_{\rm e}$=10,000--15,000 K (Fu \& Stockton 2008), which imply that they should be gravitationally unstable and have already started forming stars. Hence Fu \& Stockton (2008) modeled these regions using two components, a low density component, where most of the mass is locked at densities of order n$_{\rm e}\sim$1 cm$^{-3}$, and a higher density component responsible for the observed emission lines. Hydrodynamical simulations of radiative shock-cloud interactions indicate that for moderate gas cloud densities ($>$ 1 cm$^{-3}$), cooling processes can be highly efficient and result in more than 50\,\% of the initial cloud mass cooling to below 100 K (Fragile et al. 2004). Hence, the companion galaxy of HE0450$-$2958 could have formed from a seed EELR that was hit by the radio jet. This is the scenario that we favor here.

The following question is then what formed the seed EELR in the first place. It is not possible to infer its origin from existing datasets in the field of HE0450$-$2958. However by analogy with other systems where EELRs are found, it may be considered that the seed EELR was produced by the QSO itself in a first stage as suggested by Fu \& Stockton (2008, see also Krause \& Gaibler 2009). In the case of HE0450$-$2958, this is particularly problematic since no host galaxy has yet been detected for this QSO. However, no definitive evidence for the absence of a host galaxy has yet been demonstrated and other sources of fresh gas fuel may be considered than the interstellar medium of the host galaxy itself as discussed in Sect.~\ref{SEC:origin}.

\subsection{Extended emission line regions as potential galaxy progenitors} 
\label{SEC:EELR}
We have seen that HE0450$-$2958 was surrounded by several EELRs, as it is often the case for radio quasars. Most will not be hit by the radio jet, however the thermodynamic conditions in these gas clouds may imply that they will become the progenitors of new galaxies after the QSO stop photoionizing them. At densities of 100 cm$^{-3}$ or more, gravitational instabilities will form in a free fall time of about 80 million years (Eq.~\ref{EQ:tff}) with a Jeans mass of the order of 6$\times$10$^{6}$ M$_{\odot}$ (Eq.~\ref{EQ:MJeans}). This is the typical mass of a giant molecular cloud (GMC). If the gas was able to produce molecules, it would continue to cool down and start forming stars. As long as the quasar remains active, the gas will remain photoionized and this process will be prevented. However after the QSO will have switched off, the process may take place, then it is to be expected that stars will form. As a result, EELRs in general may be considered as seed galaxies, hence providing a potential explanation for the observed excess of companion starburst galaxies surrounding quasars (see e.g. Hutchings, Crampton \& Johnson 1995, Hutchings 1995). 

\begin{equation}
\tau_{\rm free-fall} = \frac{1}{\sqrt{G \rho}} \simeq 80 ~{\rm Myr}
\label{EQ:tff}
\end{equation}
\begin{equation}
c_{\rm sound} = \left(\gamma \frac{P}{\rho} \right)^{1/2} = \left(\gamma {T} \frac{k}{\mu} \right)^{1/2}\simeq 9 ~{\rm km} ~{\rm s}^{-1}
\label{EQ:csound}
\end{equation}
\begin{equation}
L_{\rm Jeans} = \frac{c_{\rm sound}}{\sqrt{G \rho}} = 76 ~{\rm pc}
\label{EQ:LJeans}
\end{equation}
\begin{equation}
M_{\rm Jeans} = \frac{4}{3} \pi ~L_{\rm Jeans}^3 ~\rho = 6 \times 10^6 ~ {\rm M}_{\odot}
\label{EQ:MJeans}
\end{equation}

\section{Searching for the host galaxy of HE0450$-$2958}
\label{SEC:scenario}

\subsection{An ejected supermassive black hole ?}
A possible explanation for the existence of a so-called naked quasar was developed by a series of papers claiming that such discovery could be the natural consequence of a mechanism predicted by theory during the merging of SMBH.
A dynamical kick may have been imparted to the quasar as it interacted with a binary black hole system during a galaxy merger event (Hoffman \& Loeb 2006, see also Mikkola \& Valtonen 1990) or the ejection could have been caused by gravitational radiation recoil during the coalescence of a binary BH (Haehnelt, Davies \& Rees 2006, see also Favata et al. 2004). In a recent paper, Letawe et al. (2009) interpreted the NICMOS imaging of the companion galaxy as evidence for  the presence of a strongly reddened active galactic nucleus hidden behind a thick dust cloud which would favor the ejection hypothesis during the merger of two active nuclei. Although this study also used the same VISIR data as discussed in the present paper, it was done before our refined analysis showing that both the QSO and companion galaxy are detected with VISIR. As discussed in Sect.~\ref{SEC:balance}, the companion galaxy optical to far infrared SED is compatible with a pure starburst and no evidence has yet been found for the presence of a second AGN in this system. 

For a radiation efficiency of $\eta$=0.1, the accretion rate required to produce a bolometric luminosity of L$_{\rm bol}\sim$2$\times$10$^{12}$ L$_{\odot}$ is: $\dot{M}$= L$_{\rm bol}$/($\eta$c$^2$)$\sim$1.3 M$_{\odot}$ yr$^{-1}$. Papadopoulos et al. (2008a) computed that a disc mass of M$_{\rm disc}\leq$ 2$\times$10$^6$ [M$_{\rm BH}$/(10$^7$ M$_{\odot}$)]$^{2.2}$ M$_{\odot}$ could be carried out by a recoiling black hole. For a SMBH mass of M$_{\rm BH}$$\sim$(7$\pm$2)$\times$10$^7$ M$_{\odot}$ (see Sect.~\ref{SEC:BHmass}), this would correspond to M$_{\rm disc}\leq$ 1.4$\times$10$^8$ M$_{\odot}$.  For such a disc mass, the QSO activity could last up to $\sim$100 Myr. However, the relative velocity between the quasar ($z$$=$0.2863) and its companion galaxy ($z$$=$0.2865) is only 60 km s$^{-1}$ in the line of sight and between 60 to $+$200 km s$^{-1}$ when considering emission lines in both optical spectra (Letawe et al. 2008b). This is much below the central escape velocity of the companion galaxy of more than 500 km s$^{-1}$ (Merritt et al. 2006). Such kinematical consideration alone is difficult to reconcile with the kick of the QSO out of the companion galaxy (Merritt et al. 2006). We also note that it would be statistically unexpected that the radio jet coming out of the QSO would point directly at the companion galaxy after it was ejected out of it, as it is the case here (Feain et al. 2007), since there is no obvious reason why the radio jet direction would be the same than the one of the ejection of the QSO.

To conclude on this first scenario, we believe that there is converging evidence against it but that this hypothesis cannot be definitely ruled out until a host galaxy is detected associated with the QSO. 	
\subsection{A lower mass black hole ?}
\label{SEC:BHmass}
A second possibility to explain the absence of detection of a host galaxy for HE0450$-$2958 was initially proposed by Merritt et al. (2006) and others afterwards. The reason here would be that the mass of the central SMBH was overestimated by Magain et al. (2005) and consequently also that of the bulge stellar mass predicted by the Magorrian relation for its host galaxy. This would allow the host galaxy to remain below the HST V-band detection limit and still agree with the Magorrian relation. 

A mass of $M_{\rm BH}$$\sim$8$\times$10$^8$ M$_{\odot}$ ($M_{\rm V}=-$25.8) was derived by Magain et al. (2005) assuming an accretion at half the Eddington rate and a radiation efficiency of $\eta$$=$0.1. The authors claimed that this mass agreed with the one derived from the width of the H$\beta$ line but a reanalysis of this line using VLT--FORS data  led to a lower mass determination of $M_{\rm BH}$$\sim$(9$\pm$1)$\times$10$^7$ M$_{\odot}$ by Merritt et al. (2006) and $M_{\rm BH}$$\sim$4.3$\times$10$^7$ M$_{\odot}$ by Letawe et al. (2007). Merritt et al. (2006) added that this lower SMBH estimate was also consistent with the one derived from the H$\alpha$ line of $M_{\rm BH}$$\sim$6$^{+5}_{-3}$$\times$10$^7$ M$_{\odot}$. Hence globally, the mass of the SMBH that would best fit with the observed properties of HE0450$-$2958 is of order $M_{\rm BH}$$\sim$(7$\pm$2)$\times$10$^7$ M$_{\odot}$. Adopting the $M_{\rm BH}$--NIR bulge luminosity observed locally (Marconi \& Hunt 2003) and a SMBH mass of $M_{\rm BH}\sim$(9$\pm$1)$\times$10$^7$ M$_{\odot}$, Merritt et al. (2006) inferred a $K$-band absolute magnitude for the stars in the bulge of the host galaxy of $M_K\simeq -23.4$ and a more uncertain total visual magnitude of $M_V\simeq -21$ assuming an Sa host. This visual magnitude would put the putative host galaxy marginally (by a factor 1.2 in luminosity) below the brightest V-band detection limit inferred for the HST-ACS image by Magain et al. (2005) for a 10 billion years old stellar population (-21.2$\leq M_V$). If instead, a young stellar population like the one of the companion galaxy was assumed for the SMBH host galaxy, then the host galaxy luminosity predicted by Merritt et al. (2006) would be high enough to have been detected by the HST-ACS V-band image (upper limit of -20.5$\leq M_V$ from Magain et al. 2005). We note that using a different approach than Magain et al. (2005), Kim et al. (2007) derived an upper limit for the host galaxy stellar mass of -21$\leq M_V \leq$-20 consistent with the one derived by Magain et al. (2005), which would also imply a very marginal non detection of the host galaxy for the visual magnitude inferred by Merritt et al. (2006) of $M_V\simeq -21$. 

Hence, depending on the assumed stellar population for the SMBH host galaxy, it would have to be either above the HST detection limit or only marginally below it, by a factor 1.2 in V-band luminosity. In this context, the absence of detection of a host galaxy for HE0450$-$2958 would only marginally be explained. A more direct constraint on the host galaxy mass can be derived from near IR imaging with HST-NICMOS since it is less dependant on the previous star formation history of the galaxy than the V-band and less affected by dust extinction. The QSO is detected with an apparent magnitude of $H$$\sim$13.6 (Table~\ref{TAB:photometry}) and after PSF subtraction of its emission an upper limit to the host galaxy of 3\,\% of this magnitude, hence $H$$>$16.9, was obtained by Jahnke et al. (2009). In comparison, the $K$-band absolute magnitude inferred by Merritt et al. (2006) for the host galaxy would correspond to $M_H$$\simeq$-23.3,  hence $H$$\sim$17.5 at the redshift of HE0450 (we used the SED of Arp 220 at the redshift of the quasar for the $H-K$ color correction, see Sect.~\ref{SEC:composite}). This is again (although slightly more than for the visual magnitude) marginally below the NICMOS detection limit  by 0.6 mag, or a factor 1.7 in luminosity.

To conclude, there is indeed converging evidence that the SMBH mass derived by Magain et al. (2005), and consequently the predicted stellar mass of its host galaxy, were both overestimated. It is therefore possible that HE0450$-$2958 lies inside a host galaxy that would be marginally consistent with the local Magorrian relation. 
 
\subsection{A dust-enshrouded host galaxy ?}
\label{SEC:dusty}
The observed $V$-band (6060\,\AA\,) corresponds to the $B$-band (4710\,\AA\,) where dust extinction can be very efficient. Hence, a possible explanation for the non detection of a host galaxy by Magain et al. (2005) could be the effect of dust obscuration. The VISIR image shows clearly that there is strong dust obscuration at the position of the QSO, contrary to the initial claim by Magain et al. (2005). The presence of a strong infrared emitter associated with the QSO opens the possibility for the existence of a more massive galaxy than the one which may be inferred from the HST upper limit. Yet, the dust radiation powered by the QSO may originate from a dust torus surrounding the QSO and does not necessarily imply that the putative stellar population of the host galaxy as a whole is strongly affected by dust obscuration.

In the HST-NICMOS $H$-band data (1.6\,$\mu$m observed, 1.25\,$\mu$m in the rest-frame) where the impact of dust attenuation is reduced by a factor $\sim$4.7 assuming the extinction law of the Milky Way (Fitzpatrick 1999), no evidence for a host galaxy was found brighter than 5\,\% of the QSO luminosity, which is the technical detection limit due to PSF uncertainties (Jahnke et al. 2009). The host galaxy mass expected from the local Magorrian relation would only be 1.7 times fainter than the NICMOS $H$-band detection limit, hence again only marginally below the HST detection limit. In order to increase the ratio between the NICMOS detection limit and the $H$-band magnitude of the host galaxy by a factor 5, so that it would clearly be outside of the detection range, one would have to invoke an attenuation as large as $A_V\sim$ 4 over the whole host galaxy. We also note that the H$\alpha$/H$\beta$ ratio in the optical spectrum of the QSO and the presence of photo-ionized gas clouds up to 30 kpc away from the QSO indicate that the line of sight to the QSO nucleus is nearly dust-free (Letawe et al. 2008a).

To conclude, the ULIRG status of the QSO HE0450$-$2958 does provide evidence that strong dust obscuration is taking place in this object but not necessarily at the scale of a whole galaxy. Hence the presence of a dust-obscured host galaxy cannot be rejected but no definitive evidence for its presence exists at this stage.

   \begin{figure*}[ht!]
   \centering
   \includegraphics[width=8cm]{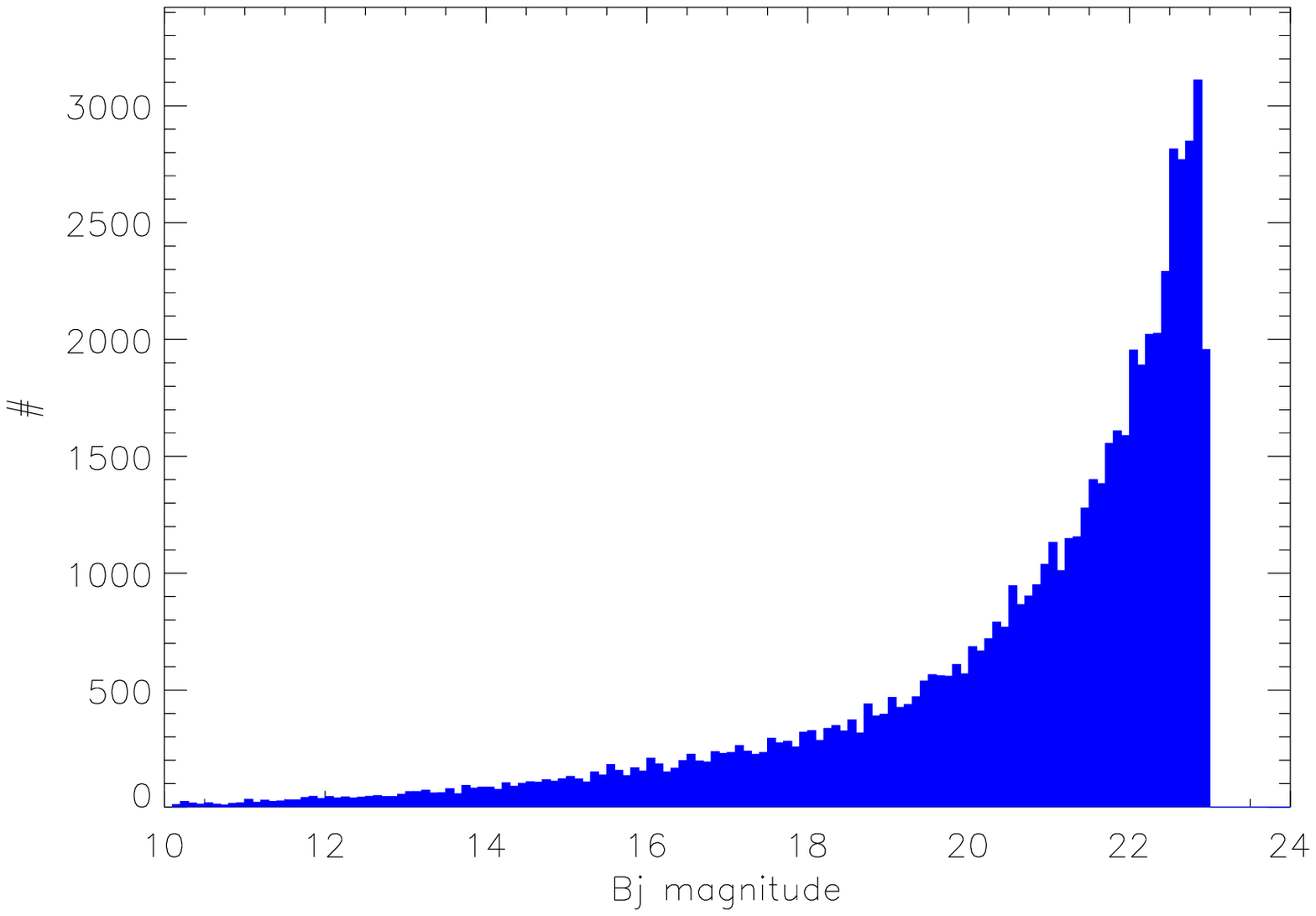}
      \includegraphics[width=8cm]{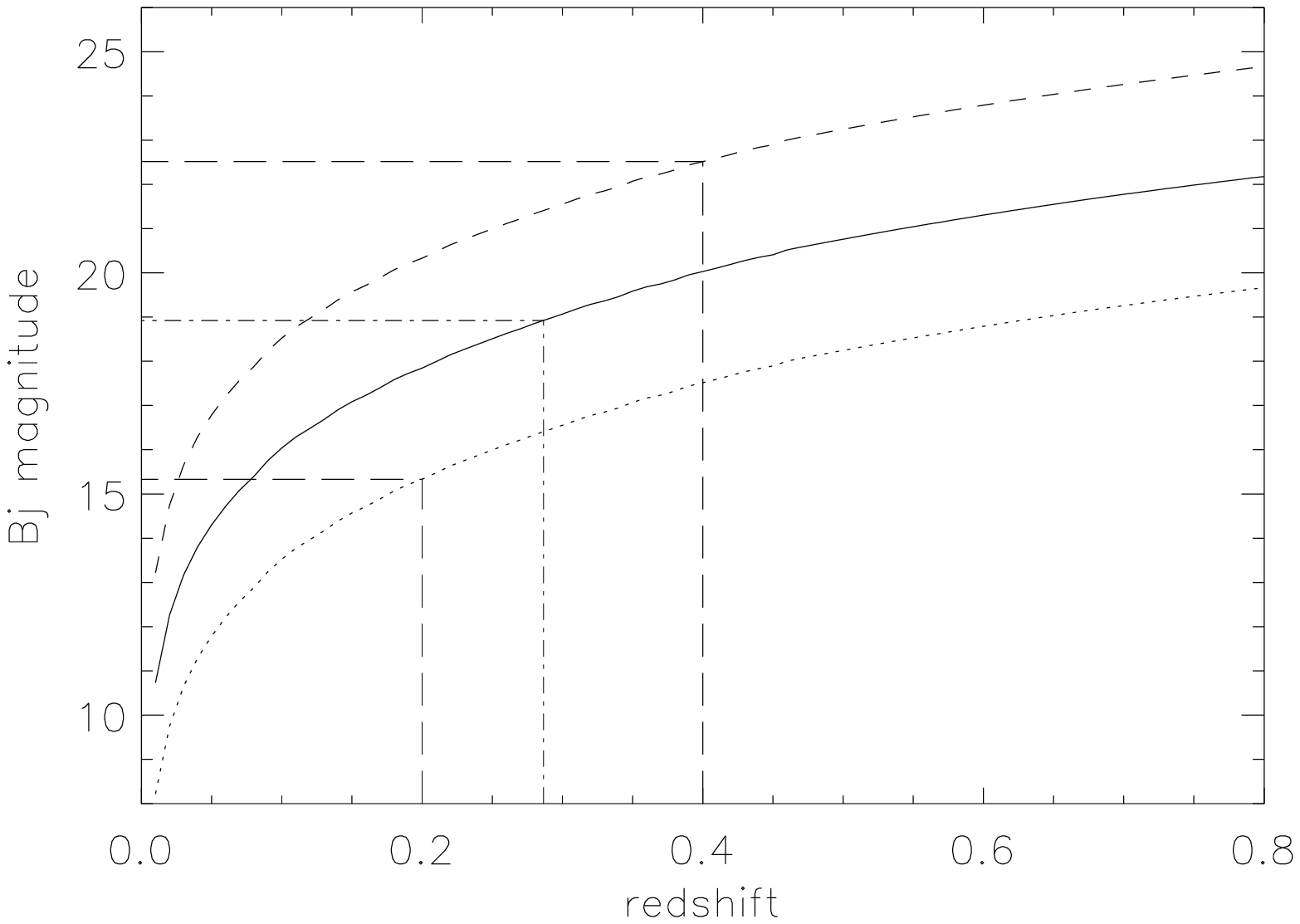}
      \caption{\textbf{\textit{(left)}} Distribution of the $b_J$ apparent magnitudes in the Vega system (3900A-5400\,\AA\,passband centered at 4604\AA\, with a zero point of F($b_J$, Vega)= 3.95$\times$10$^{23}$ W m$^{-2}$ Hz$^{-1}$) of the galaxies located in a region of 2.5\degr\, on a side centered close to the position of HE0450$-$2958 from the APM galaxy survey (Maddox et al. 1990). The detection limit of the catalog is around $b_J$=22.5 (stars excluded from the catalog). \textbf{\textit{(right)}} Projected evolution with redshift of the $b_J$ apparent magnitude of the companion galaxy of HE0450$-$2958 (plain line) as well as a ten times less luminous (dashed curve) or more luminous (dotted curve) galaxy. The dash-dotted horizontal and vertical lines indicate the actual position of the galaxy in this diagram. The horizontal/vertical long-dashed lines mark the redshift range of $z$=0.2--0.4 in this luminosity range of 0.1 and 10 times the companion galaxy bolometric optical luminosity of $L_{\rm bol}^{\rm opt}$=3$\times$10$^{11}$ L$_{\sun}$.
              }
         \label{FIG:magsBj}
   \end{figure*}

\subsection{Is the companion galaxy of HE0450$-$2958 its future host galaxy ?}
\label{SEC:hostform}
We have reviewed a series of possible explanations for the absence of detection of a host galaxy for HE0450$-$2958. Revisions of the SMBH mass make the case of HE0450$-$2958 potentially "normal" with respect to the local Magorrian relation, although only marginally (by less than a factor 2). Adding some level of extinction would make the host galaxy more comfortably below the ACS and NICMOS detection limits but we present in Sect.~\ref{SEC:dusty} evidence against a highly obscured host galaxy. 

Considering the quasar (M$_{\rm BH}$$\simeq$[7$\pm$2]$\times$10$^7$ M$_{\odot}$, see Sect.~\ref{SEC:BHmass}) and its companion galaxy (M$_{\star}$$\simeq$[5--6]$\times$10$^{10}$ M$_{\odot}$, Sect.~\ref{SEC:stars}) as a single system, it would exhibit a standard M$_{\star}^{\rm bulge}/$M$_{\rm BH}$ ratio of $\simeq$ 780 when compared to the typical one found for local galaxies of about 500 (Marconi \& Hunt 2003, McLure \& Dunlop 2001, Ferrarese et al. 2006), 700 (Kormendy \& Gebhardt 2001) or 830 (McLure \& Dunlop 2002). The relative velocity of the quasar ($z$$=$0.2863) and companion galaxy ($z$$=$0.2865) in the line of sight is only 60 km s$^{-1}$. Using the emission lines in their spectra, Letawe et al. (2008b) derived relative velocities ranging from 60 to $+$200 km s$^{-1}$. Hence, e.g. if the system belongs to a common dark matter halo, it is nearly unavoidable that with such small relative velocity, the quasar and its companion galaxy -- distant of only 7 kpc or 1.5\arcsec\,-- will ultimately merge together (see also Jahnke et al. 2009). This would provide a natural explanation for the missing host galaxy. 

Hence if HE0450$-$2958 and its companion galaxy were to end up merging together, as it is nearly unavoidable, the system would fall on the Magorrian relation and there would remain no signature that the major fraction of the stars located in the final host galaxy formed outside the region occupied by the original QSO. In this framework, HE0450$-$2958 would be an exceptionally young quasar in the process of building its host galaxy. If it is later on found from deeper near infrared imaging that there already exists an underlying host galaxy centered on the QSO, its merging with the companion galaxy would remain highly probable. In both cases, the star formation event that is taking place in the companion galaxy is highly informative of the way the Magorrian relation is built for this galaxy. If not unique to this system, these observations imply that jet-induced star formation plays a key role in the origin of the physical connection between host galaxies stellar mass and the building of supermassive black holes.

\section{A possible origin for the quasar activity of HE0450$-$2958: accretion from intergalactic cold gas filaments}
\label{SEC:origin}
The question of the origin of the quasar activity is not specific to HE0450$-$2958, since even in systems with a well detected host galaxy the physical mechanism that would bring large  amounts of matter at the sub-parsec scale of the central SMBH remains highly uncertain (King \& Pringle 2007, Thompson, Quataert \& Murray 2005). On larger scales mergers have been advocated to solve the problem of the necessity for the ISM of the host galaxy to loose its angular momentum, but this remains a matter of debate. Indeed, looking at the statistical properties of AGNs in the SDSS, Li et al. (2008) did not find any evidence for a connection of AGN activity with galaxy pairs.

In Sect.~\ref{SEC:scenario}, we considered various possible explanations for the absence of detection of a host galaxy for HE0450$-$2958. 
We proposed in Sect.~\ref{SEC:hostform} that the companion galaxy of HE0450$-$2958 could be its whole future host galaxy or that it will, at least, participate to its mass building since both objects are bound to ultimately merge together. The gas blob adjacent to the QSO in the N-W direction may be considered as a potential source of fuel for the QSO. However, it is located in the direction of the radio jet opposite to the companion galaxy and we detect some mid-infrared emission associated with it as well as on the opposite side of the QSO (see Fig.~\ref{FIG:contours}a), which may result from matter expelled together with the radio jets of the QSO. As a result we do not consider the gas blob as a strong candidate for material fueling the QSO.

We now discuss the issue of the physical origin of the quasar activity of this object. It is generally assumed that the source of material feeding a QSO comes from the interstellar matter of its host galaxy but even in the presence of a host galaxy, the physical mechanism that would bring large amounts of matter at the scale of the central SMBH remains highly uncertain (see e.g. Thompson, Quataert \& Murray 2005). 

The vast majority of baryons (90\,\%) are located outside galaxies and for a large fraction of them, in the form of intergalactic filaments. It is well-known that in galaxy clusters, intracluster gas can cool and fall onto the central dominant (cD) galaxy participating to the activation of its central AGN activity and providing fuel for very large outflows reaching up to 1000 M$_{\odot}$ yr$^{-1}$ (Nesvadba et al. 2006, 2008). In the more distant Universe, this must have been even more common. It has been recently suggested that the bulk of the baryonic mass of galaxies might have been gathered through the accretion of intergalactic gas filaments instead of mergers (Dekel et al. 2009). We note that these filaments are not necessarily composed by a continuous and homogeneous gas distribution but may instead be clumpy
as inferred from these simulations. If this mechanism is indeed an important one that feeds star formation in galaxies with large amounts of fresh material, then it may also provide some fuel for active galactic nuclei and quasars (see also Park \& Lee 2009). In that case, QSOs might be maintained active without needing to be located inside a gas-rich host galaxy and HE0450$-$2958 might be a local illustration of such process at an early stage. 

In order to search for intergalactic gas filaments in the area of HE0450$-$2958, we map the projected large-scale distribution of individual galaxies within a region of 2.5\degr\,on the side, corresponding to a comoving size of 52 Mpc at the redshift of $z$=0.2863. Since no complete spectroscopic redshift survey of the field exists at present, we study the projected environment of the QSO and select slices of the Universe at various putative depths in the redshift direction from a magnitude selection. The best imaging dataset of this field of view comes from the APM galaxy survey (Maddox et al. 1990), a galaxy catalog digitized from scan plates taken at the UK Schmidt Telescope Unit (UKTSU) using the SERC Automatic Plate Measuring facility in Cambridge. We also retrieved galaxies from the NASA Extragalactic Database (NED) including sources from 2MASS, IRAS and Abell galaxy clusters. The closest IRAS galaxy to HE0450$-$2958 is 18\arcmin\,distant. Two galaxy clusters are present in a region of 2.5\degr$^{2}$ centered on the QSO and the closest galaxy cluster is 25\arcmin\,distant.

   \begin{figure}
   \centering
\includegraphics[width=9cm]{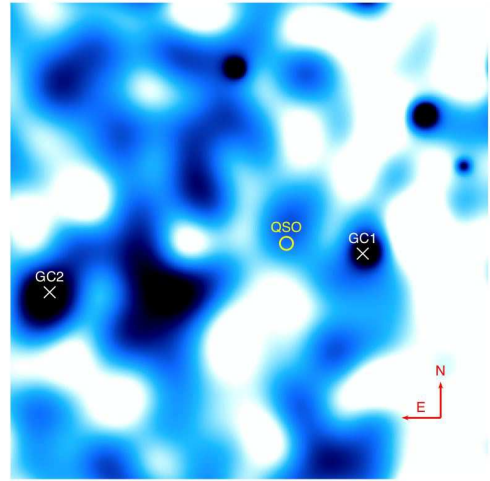}
      \caption{Density map of the projected distribution of galaxies in the field of view of HE0450$-$2958 (red circle in the middle of the image). Darker regions indicate denser regions as inferred from the wavelet filtering decomposition of the projected spatial distribution of the 60,000 galaxies detected by the APM survey in a field of 2.5\degr\,on the side. This size would correspond to a comoving size of 52 Mpc if all galaxies were located at the redshift of $z$=0.2863 of HE0450$-$2958. GC1 and GC2 are two galaxy clusters located in this field. Here we used the full range of apparent magnitudes from $b_J$=15 to 22.5.
              }
         \label{FIG:clusters}
   \end{figure}
A total of about 60,000 galaxies were detected in this region by the APM survey down to a limiting apparent magnitude of $b_J$=22.5 (see Fig.~\ref{FIG:magsBj}). The optical magnitude of the companion galaxy of HE0450$-$2958 in this passband would be $b_J$=18.9 using the model fit presented in Sect.~\ref{SEC:stars}. We use this galaxy as a reference to study the putative redshift distribution of the APM galaxies. The plain line in Fig.~\ref{FIG:magsBj}-right shows the $b_J$ apparent magnitude that the companion galaxy would have if observed at a redshift ranging from to $z$=0.01 to 0.8. The dash-dotted horizontal and vertical lines indicate the actual position of the galaxy in this diagram. The bolometric optical luminosity of the companion galaxy (excluding the mid to far infrared part of the spectrum) is $L_{\rm bol}^{\rm opt}$=3$\times$10$^{11}$ L$_{\sun}$. We show the range of apparent magnitudes that galaxies with ten times lower (dashed line) and higher (dotted line) optical luminosities would exhibit in the same redshift range. One can see that the redshift range $z$=0.2--0.4 surrounding the actual redshift of the companion galaxy ($z$=0.2865) is well covered by selecting sources in the magnitude range $b_J$=15 to 22.5.

   \begin{figure*}[ht!]
   \centering
   \includegraphics[width=9cm]{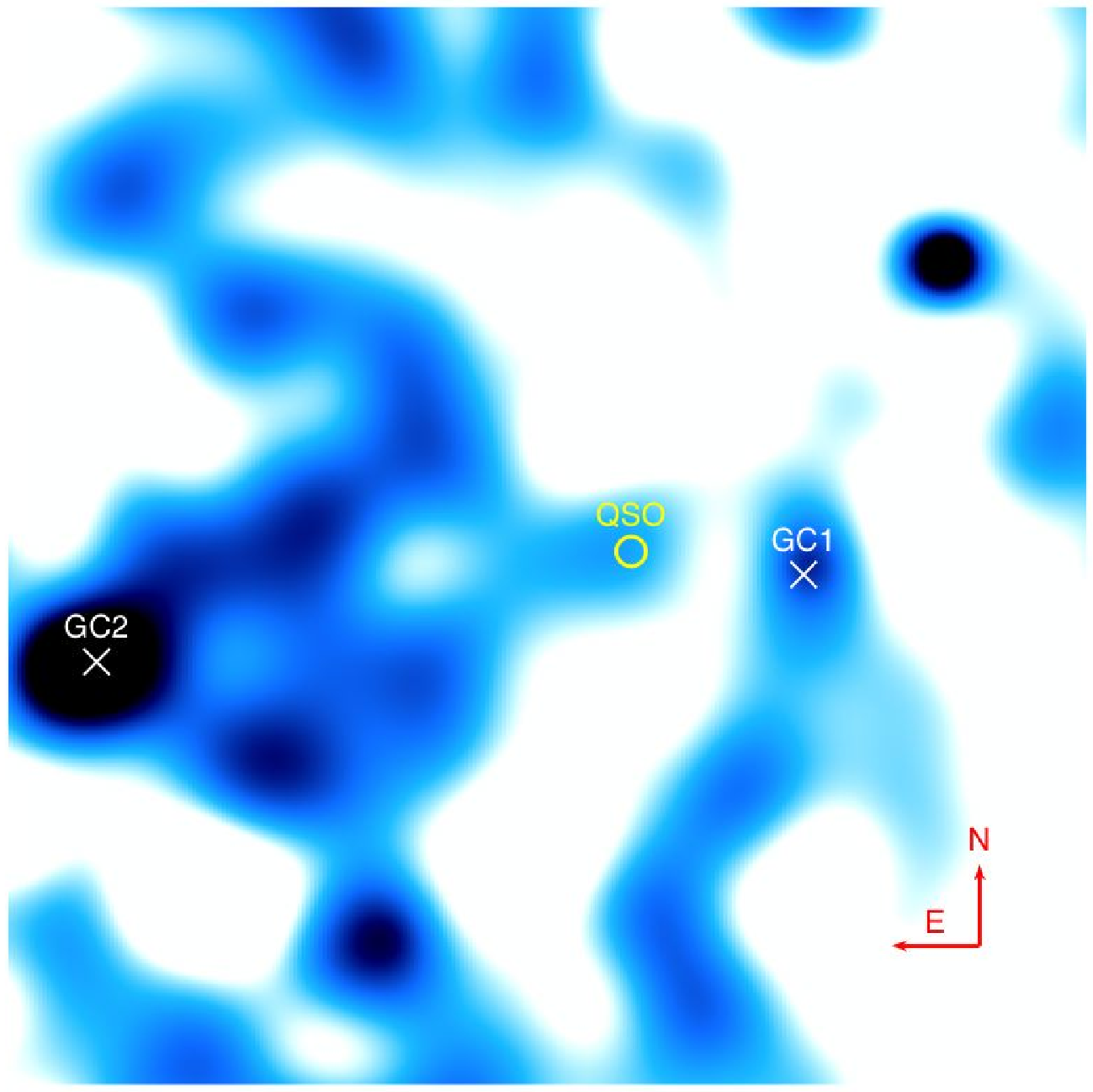}
   \includegraphics[width=9cm]{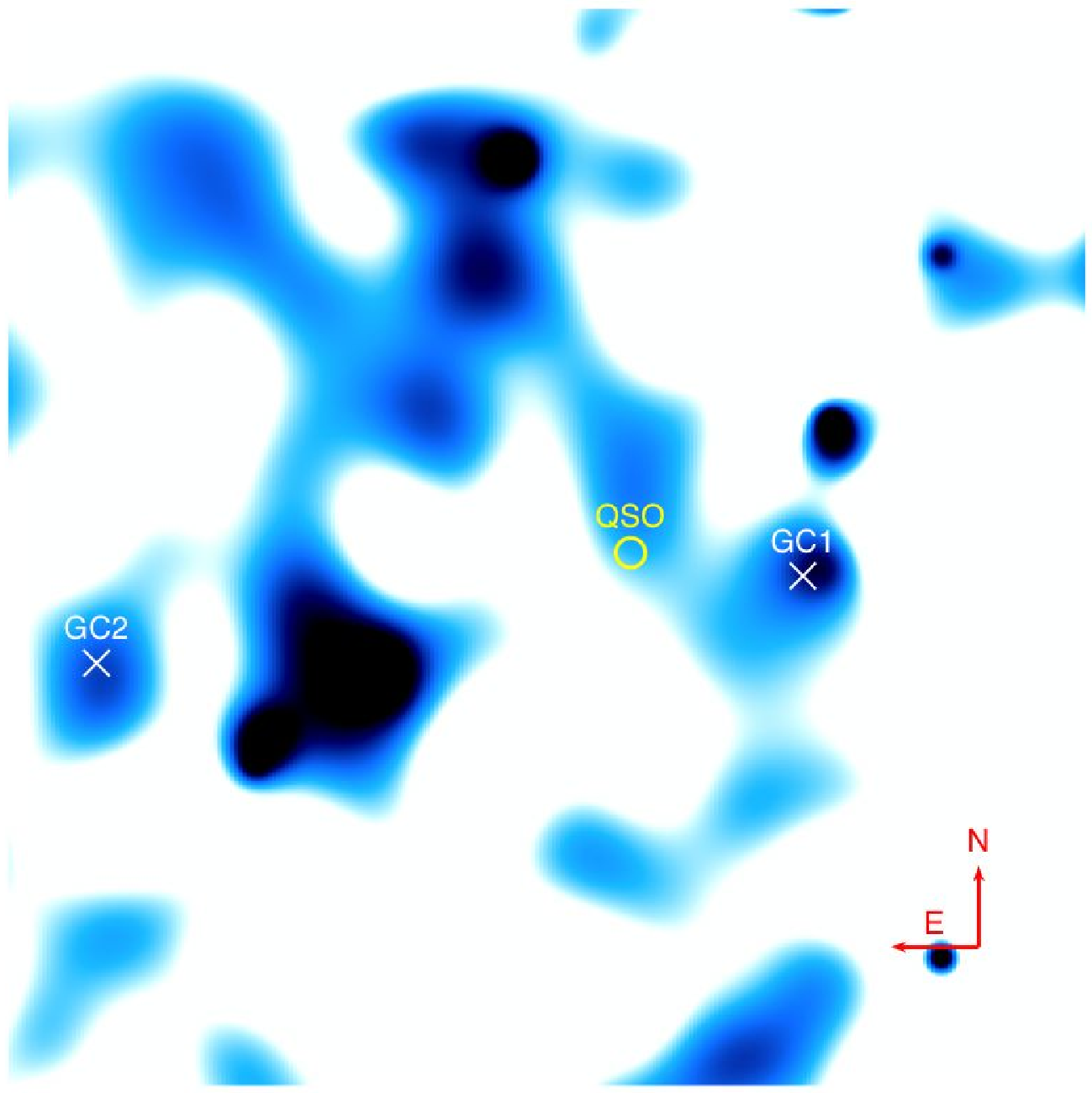}
      \caption{Density maps of the projected distribution of galaxies in the field of view of HE0450$-$2958 (red circle in the middle of the images) as in Fig.~\ref{FIG:clusters} after separating galaxies in two intervals of apparent magnitude. \textbf{\textit{(left)}} bright galaxies (20.5$\geq$$b_J$$\geq$15.5), hence also statistically closer objects, i.e. in the foreground of HE0450$-$2958 and supposedly closer to the redshift of the galaxy cluster with a determined redshift, GC2 or Abell 3297 ($z$=0.107, David, Forman \& Jones 1999). \textbf{\textit{(right)}} map of the more distant density field (22.5$\geq$$b_J$$>$20.5) showing the presence of a filamentary structure in the galaxy distribution possibly associated with the QSO.
              }
         \label{FIG:density}
   \end{figure*}
We applied a multi-resolution wavelet decomposition of the galaxy distribution (as in Elbaz et al. 2007) using seven wavelet scales (scale 1 has a 35 comoving-kpc size, while scale 7 has a 35 kpc$\times$2$^6$ = 2.2 comoving-Mpc size) to produce projected galaxy density contours. In the density map produced using all galaxies within the magnitude range $b_J$=15 to 22.5 several density peaks are found, including one at the position of HE0450$-$2958 (Fig.~\ref{FIG:clusters}). Two peaks are found associated with previously known galaxy clusters that we labeled GC1 and GC2. GC2 is an Abell cluster, Abell 3297 (4h58m29.5s, -30\degr\,08\arcmin\,31\arcsec), for which a redshift of $z$=0.106 was determined from the apparent magnitude versus redshift relation of Ebeling et al. (1996) by David, Forman \& Jones (1999). GC1 was identified as a cluster by an automatic cluster search in the APM galaxy catalog and classified as APMCC539 (4h50m40s, -29\degr\,56.7\arcmin). No redshift was determined for this cluster. One of the sources in its area has a redshift of $z$=0.1076 from 2MASS (2MASX J045755-3009380), hence similar to the one of Abell 3297.

In order to separate foreground galaxies from galaxies in the local environment of HE0450$-$2958, we divided galaxies in two bins of apparent magnitudes, above and below $b_J$= 20.5. The resulting density maps are shown in Fig.~\ref{FIG:density}, with the bright galaxy distribution ($b_J$$\leq$ 20.5) on the left and fainter galaxies ($b_J$$>$ 20.5) on the right. The most proeminent structure in the closest galaxy population (Fig.~\ref{FIG:density}-left) is GC2, i.e. Abell 3297, at $z\sim$0.1. This structures becomes marginal in the fainter sample of expectedly more distant galaxies (Fig.~\ref{FIG:density}-right). Instead, we note that GC1 is slightly more contrasted in the faint galaxy sample, which may suggest that it is located further away than GC2. A spectroscopic follow-up of some of its galaxies would be needed to determine whether GC1 is physically close to HE0450$-$2958. If GC1 was located at the redshift of HE0450$-$2958, the QSO would be located at a projected distance of 8 comoving-Mpc from the cluster center (25\arcmin). The separation of both galaxy populations resulted in putting more emphasis on a linear distribution of galaxies along what may be a large-scale filament going through the position of the QSO. It is difficult to conclude on the basis of apparent magnitudes only that this filamentary structure in the density map is both real and physically linked to HE0450$-$2958, but it is an interesting candidate galaxy filament which might be tracing a large-scale structure following which gas infall might be taking place. 

The role of intergalactic gas filaments has not yet been considered as potentially important in the history of SMBH growth but if SMBH grew faster than their host galaxies, then new mechanisms might need to be considered in order to explain the source of material feeding the central object. We searched for alternative scenarios in the literature and the only one that we were able to find was published in the late 90's, when it was suggested that quasars may be produced as a result of superconductive cosmic string loops (Vilenkin \& Field 1987, Manka \& Bednarek 1991). Cosmic strings are linear defects that could be formed at a symmetry breaking phase transition in the early Universe (Vilenkin \& Shellard 1994). Once formed, cosmic strings would exist at any time in a network of loops and inÞnite strings. These loops might radiate away their content in what those authors considered to be an event that would look alike observed quasars with radio jets. Interestingly, a loop of R=10 kpc, the typical size of a galaxy, is expected to contain a mass of 6$\times$10$^{11}$ M$_{\odot}$, the typical mass of a galaxy, equivalent to the stellar mass of the Milky Way. 

\section{Discussion on the role of quasars in galaxy formation}
\label{SEC:roleQSO}
We presented observations suggesting that the companion galaxy of HE0450$-$2958 was recently formed from the shock of radio jets on a seed EELR. In this scenario, the QSO would switch on first and act as a triggering mechanism for the rapid star formation of the future bulge. We also discussed the possibility that EELRs, in general, might end up forming stars, after their gas stopped being photoionized by the QSO, when the QSO activity ends. Indeed, their dense parts ($>$ 100 cm$^{-3}$) are in the thermodynamical conditions to form stars after the quasar stops photoionizing their gas (Sect.~\ref{SEC:form_companion}). A theoretical discussion on how quasar outflows may generate dwarf galaxies can also be found in Natarajan, Sigurdsson \& Silk (1998). Here we wish to list a series of observations that provide converging evidence that quasars play a key role in the formation of galaxies:
\begin{enumerate}
\item The local M$_{\rm BH}$--M$_{\star}^{\rm bulge}$ correlation suggests that there is a physical mechanism connecting star formation and AGN activity in galaxies.
\item Evidence for jet-induced star formation in radio galaxies/QSOs suggests that radio jets might represent such physical mechanism.
\item At least two systems, Minkowski's object and now HE0450$-$2958, might be representative of an early stage of quasar induced galaxy formation.
\item The radio-optical alignment of some QSO host galaxies might be representative of a later stage of quasar induced galaxy formation, a fossil memory of the process through which star formation was triggered inside the QSO host galaxy.
\item Extended emission line regions (EELRs), often found around radio quasars, may be considered as the candidate seeds for future galaxy formation.
\item Several distant QSOs present evidence for an associated large molecular gas mass shifted in position with respect to the QSO itself, suggesting a link between the QSO and CO concentration through the QSO radio jets.
\item Finally, if QSOs formed first and their host galaxy in a second stage -- or at least the final bulge stellar mass took longer to be built -- then one may expect to measure lower M$_{\star}^{\rm bulge}/$M$_{\rm BH}$ ratios in distant QSOs. Several studies discussed in Sect.\ref{SEC:highz} have recently claimed to have found evidence of such trend.
\end{enumerate}

In the following, we discuss some of these observations in more detail. The first four points of the list, concerning the role of radio jets, are discussed in Sect.~\ref{SEC:jetinduced}, point 5 was discussed in Sect.~\ref{SEC:form_companion}, point 6 is discussed in Sect.~\ref{SEC:COoffset} and point 7 in Sect.~\ref{SEC:highz}.

We note that jet-induced star formation may provide an alternative explanation to mergers for the very short star formation timescale of bulge stars (e.g. Thomas et al. 2005, Pipino, Silk \& Matteucci 2009). It may also participate to the origin of the downsizing effect (Cowie et al. 1996), which refers to the anticorrelation between the stellar mass and the formation epoch of the stars of galaxies (Cowie et al. 1996, Guzman et al. 1997, Brinchmann \& Ellis 2000, Kodama et al. 2004, Heavens et al. 2004, Bell et al. 2005, Bundy et al. 2005, Juneau et al. 2005, Thomas et al. 2005, Panter et al. 2007), by accelerating the formation of massive galaxies which would preferentially host more active quasars. Hence the downsizing in star formation would be a consequence of the downsizing of active nuclei (see e.g. Hasinger, Miyaji \& Schmidt 2005).

\subsection{The role of radio jets in star and galaxy formation} 
\label{SEC:jetinduced}
 The role of radio jets in star formation remains a matter of debate and has been oscillating between two apparently opposite interpretations, the negative and positive feedbacks either quenching or triggering star formation in galaxies.

Evidence for jet-induced star formation has been found in various objects and environments, either far away from the host galaxy, such as in the lobe of radio jets (e.g. van Breugel et al. 1985), or inside the host galaxy, resulting in the so-called radio-optical alignment (e.g. McCarthy et al. 1987, Rees 1989). McCarthy et al. (1987) noticed that both the stellar continuum and size of the emission line region of 3CR radio galaxies at $z\ge$0.6 were highly elongated in parallel with their radio jets. This so-called radio--optical alignment was interpreted as evidence that the radio jets interact with the interstellar medium and stimulate large-scale star formation in the host galaxy (see also Rees 1989, Rejkuba et al. 2002, Oosterloo \& Morganti 2005).  Other mechanisms than jet-induced star formation have been suggested in the literature to explain the radio-optical alignment effect such as the scattering of light from the central AGN (e.g. Dey \& Spinrad 1996) or the nebular continuum emission from warm line-emitting regions (Dickson et al. 1995). However, many of these objects show clear evidence of star formation. This is, in particular, the case of 4C 41.17 ($z$=3.8), which rest-frame UV continuum emission is aligned with the radio axis of the galaxy, unpolarized and showing P Cygni-like features similar to those seen in star-forming galaxies (Dey et al. 1997). The most dramatic events, with star formation rates as high as 1000 M$_{\odot}$ yr$^{-1}$, are associated with very luminous radio galaxies at redshifts up to z $\sim$ 4 often located at the center of galaxy clusters (Dey et al. 1997, Bicknell et al. 2000, Zirm et al. 2005).

\subsection{Offset molecular gas and QSO jets}
\label{SEC:COoffset}
In HE0450$-$2958, a large mass of molecular gas is found offset with respect to the QSO, spatially associated to the companion galaxy and with the S-E radio jet (Papadopoulos et al. 2008a). The concentration of CO emission is coincident with the VISIR image of the companion galaxy, hence confirming that it is indeed the fuel of the intense starburst event. 
Such offset between molecular gas and QSOs may be a common feature in distant radio sources. Klamer et al. (2004) produced a systematic search in $z$$>$3 CO emitters for an AGN offset with respect to the CO concentrations and for a connection with radio jets. Out of the 12 $z>$ 3 CO emitters that they studied, six showed evidence of jets aligned with either the CO or dust, five have radio luminosities above 10$^{27}$ W Hz$^{-1}$ and are clearly AGNs, and a further four have radio luminosities above 10$^{25}$ W Hz$^{-1}$ indicating either extreme starbursts or possible AGNs. In the following, we discuss three proto-typical cases classified with increasing redshift from $z$$=$2.6 to 6.4 that may be compared with HE0450$-$2958.

At $z$=1.574, 3C18 is radio loud quasar with recently formed radio jets, as inferred from their small physical size. A large mass of molecular gas ($M_{\rm H_2}$$\sim$(3.0$\pm$0.6)$\times$10$^{10}$ M$_{\sun}$), inferred from CO 2--1 emission, is found associated to the quasar with a positional offset of $\sim$20 kpc (Willott et al. 2007).

At $z$= 2.6, TXS0828$+$193 is a radio galaxy with a neighboring CO gas concentration of $\sim$1.4$\times$10$^{10}$ M$_{\sun}$ located 80 kpc away and no evidence for an underlying presence of stars or galaxy (Nesvadba et al. 2009). An upper limit of 0.1 mJy at 24\,$\mu$m, from the MIPS camera onboard Spitzer, was obtained by Nesvadba et al. (2009), who concluded from this limit that no major starburst with more than several hundred solar masses per year could be taking place associated with this CO concentration. We used the library of SED templates of local galaxies from Chary \& Elbaz (2001) to convert this mid infrared measurement into an upper limit for the total IR luminosity at this position of L$_{\rm IR}^{\rm max}$$\sim$7.8$\times$10$^{12}$ L$_{\sun}$, which would translate to a maximum SFR of $\sim$1340 M$_{\odot}$ yr$^{-1}$ using the Kennicutt (1998) conversion factor for a Salpeter IMF. Hence there is still room for a large amount of star formation in this object that may be in a similar stage than HE0450$-$2958 but at a much larger distance. We note that metallicities up to nearly solar were found in the ionized gas detected in the outer halo of the galaxy suggesting that matter was driven out of the central host galaxy (Villar-Mart\'in et al. 2002).

At $z$= 4.695, BR 1202$-$0725 is the first high-redshift quasar for which large amounts of molecular gas was detected (Omont et al. 1996; Ohta et al. 1996). 
The CO map presents two well separated emission peaks which coincide with radio continuum emission interpreted as evidence for the presence of radio jets by Carilli et al. (2002). The radio jets themselves would be too faint to be detected at the sensitivity level of the radio image but marginal evidence for variability suggest that the radio emission is not due to star formation. Omont et al. (1996) suggested that the double CO emission might be due to gravitational lensing, but no optical counterpart of the QSO is found associated to the second source and the CO (2--1) line profiles are different for the two components (Carilli et al. 2002). Klamer et al. (2004) suggested that stars may have formed along the radio jets, providing both the metals and dust for cooling and "conventional" star formation to take place afterwards.

In the case of HE0450$-$2958, the mid infrared emission is spatially associated with the offset CO concentration suggesting that similar systems might be found not only through CO imaging but also infrared imaging. SDSS160705+533558, located at a redshift of $z$=3.65, might be a distant analog of HE0450$-$2958 in that respect since it also presents a positional offset between the maximum sub-millimeter (from the submillimeter arra, SMA) and optical emission (Clements et al. 2009). We note that shocks due to radio jets can induce large high-J CO line luminosities (Papadopoulos et al. 2008b).

A key test for the role of radio jets in the formation of galaxies will be the detection of either molecular gas or mid infrared emission with a positional offset with respect to their neighboring QSO. This test will be fulfilled with the advent of ALMA, the JWST or project instruments such as METIS (mid infrared E-ELT Imager and Spectrograph) for the ELT (extremely large telescope).

\subsection{The building of the M$_{\rm BH}$ - M$_{\star}^{\rm bulge}$ relation: did SMBH form first ?}
\label{SEC:highz}
If not unique, the case of HE0450$-$2958 suggests that local supermassive black hole host galaxies, or at least a large fraction of their stars, were formed outside quasars and that their star formation history was strongly affected by radio jets. The fact that at $z$$\sim$2, the comoving space density of radio galaxies with powerful jets was 1000 times higher than at the present epoch (Willott et al. 2001) suggests that radio jets may indeed play a key role in galaxy evolution, although their lower luminosity counterparts have only been studied up to $z\sim$1. This scenario would provide a natural explanation for the M$_{\rm BH}$--M$_{\star}^{\rm bulge}$ relation and it may be tested by the relative timing of formation of SMBH and host galaxies. There is already converging evidence that the M$_{\star}^{\rm bulge}/$M$_{\rm BH}$ ratio was lower in the distant Universe than locally, i.e. that SMBH may have formed first. However, due to the difficulty to estimate black hole masses in distant galaxies, this trend would have to be confirmed and sub-millimeter galaxies appear to exhibit an opposite trend (see e.g. Alexander et al. 2008), but if this trend was confirmed, this would give credit to the above scenario. 

The host galaxies of seyfert nuclei at $z$$\sim$0.37 (Treu et al. 2004) and 0.36 (Treu et al. 2007) were found to exhibit lower velocity dispersions, hence also bulge stellar masses, than expected by the local $M_{\rm BH}$--M$_{\star}^{\rm bulge}$ relation with a systematic offset implying that, assuming no significant black hole growth, the distant spheroids had to grow their stellar mass by approximately 60\,\% during the last the 4 billions years. In the redshift range 0$<$$z$$<$2, 3C RR quasars exhibit an evolution as M$_{\star}^{\rm bulge}/$M$_{\rm BH}$$\propto$(1+$z$)$^{-(2.07\pm0.76)}$, reaching M$_{\star}^{\rm bulge}/$M$_{\rm BH}$$\sim$125 at $z$=2 (McLure et al. 2006), hence 5 to 6 times lower than in the local Universe. In a study of 31 gravitationally lensed AGNs and 20 nonlensed AGNs, Peng et al. (2006) found that M$_{\star}^{\rm bulge}/$M$_{\rm BH}$ was more than four times lower at $z$$>$1.7 than it is today. By the redshift 1$\leq$$z$$\leq$1.7, they found this ratio to be at most twice lower than today. They conclude that scenarios in which moderately luminous quasar hosts at $z$$\gtrsim$1.7 were fully formed bulges that passively faded to the present epoch are ruled out. In the more distant Universe, Shields et al. (2006) found that at redshifts $z$$>$3, the CO line widths are narrower than expected for their associated black hole mass indicating that SMBH reside in undersized bulges by an order of magnitude or more. Finally, ratios as low as M$_{\star}^{\rm bulge}/$M$_{\rm BH}$$\sim$ 3.4, 30 and 10--50 were estimated for APM 08279+5255 ($z$= 3.911, Riechers et al. 2009), PSS J2322+1944 ($z$=4.12, Riechers et al. 2008) and SDSS J1148+5251 ($z$=6.42, Walter et al. 2004) respectively. 

Hence, there is converging evidence that SMBH formed first followed by the stellar mass of their host galaxy. This is at odds with the classical interpretation that mergers first trigger star formation then feed the central SMBH (e.g. Sanders et al. 1988). If SMBH were already in place at the early formation stages of what will become massive bulges, these SMBH may have actively participated to the stellar mass building of these bulges. But this evolution of M$_{\star}^{\rm bulge}/$M$_{\rm BH}$ ratio with lookback time remains to be confirmed against selection effects, uncertain measurements of BH masses in case of non virialization of the systems and uncertain stellar mass measurements.
\section{Conclusions}
\label{SEC:conclusion}
We have placed a direct observational constraint on the spatial distribution of the mid infrared emission (11.3\,$\mu$m observed, 8.9\,$\mu$m rest-frame) in the field of view of the QSO HE0450$-$2958 ($z$=0.2863). When compared to optical, near-infrared, CO and radio continuum images from the HST-ACS, NICMOS and ATCA, it is found that the second source detected by VISIR is coincident with the companion galaxy of the QSO HE0450$-$2958, which exhibits a very disturbed optical morphology, is located at the peak concentration of molecular gas as traced by the CO molecule, which avoids the QSO itself (Papadopoulos et al. 2008a) and is aligned with one of the radio jets coming out of the QSO (Feain et al. 2007). We modeled the mid to far infrared spectral energy distributions of the two sources as well as the VLT-FORS optical spectrum of the stellar emission coming from the companion galaxy and found that:
\begin{enumerate}
\item The QSO and its 7 kpc distant companion galaxy both belong to the class of ULIRGs with nearly equal total infrared luminosities of L$_{\rm IR}$=L[8--1000\,$\mu$m]$\simeq$2$\times$10$^{12}$ L$_{\sun}$. Hence HE0450$-$2958 is a composite system where dust heated by an AGN and young stars are spatially separated.
\item We derive an age for the stellar population of the companion galaxy of about 40--200 Myr suggesting that the galaxy was recently born. This timescale times the SFR derived from its IR luminosity (SFR$\sim$340 M$_{\sun}$yr$^{-1}$) is consistent with its stellar mass (M$_{\star}$$\simeq$[5--6]$\times$10$^{10}$ M$_{\sun}$).
\item We propose that the starburst taking place in the companion galaxy is jet-induced. This statement results from the spatial association of the companion galaxy with the peak radio continuum associated with one of the two radio jets produced by the QSO HE0450$-$2958, its large [NII]/H$\alpha$ emission line ratio and the presence of a gradient of ionization in the ISM of the companion galaxy following the radio jet.
\item We suggest that we are witnessing a "quasar induced galaxy formation" process following two steps: the production of concentrations of ionized gas (extended emission line regions, EELRs) ejected by the QSO followed by the impact of the radio jet in one of them that will induce star formation.
\item We present evidence suggesting that the QSO will ultimately merge with its companion galaxy. Even if a host galaxy is later on detected below present detection limits, this event appears as a major one in the building of the stellar content of the host galaxy of HE0450$-$2958. This offers a physical mechanism to explain the local M$_{\rm BH}$--M$_{\star}^{\rm bulge}$ correlation in which jet-induced star formation is responsible for a major phase in the stellar mass building of host galaxies.
\item Finally, we propose a mechanism to explain the origin of the large quantity of material required to fuel the QSO: the accretion of intergalactic cold gas filaments. 
We produce projected density maps of the 50 Mpc (comoving) environment of HE0450$-$2958 in three magnitude ranges in order to separate foreground objects and find a candidate large-scale filament in the distribution of galaxies that may be physically associated with the QSO.
\end{enumerate}

HE0450$-$2958 may be considered as a test case for the role of radio jets in galaxy formation. If host galaxies, or part of their stellar mass, were formed outside quasars and their star formation strongly affected by the quasar activity, the existence of a correlation between their bulge stellar mass and SMBH mass would be naturally explained. The idea that quasars play a key role in the formation of galaxies may be inferred from the following list of observational evidence: 

\begin{itemize}
\item the local M$_{\rm BH}$--M$_{\star}^{\rm bulge}$ correlation.
\item observational evidence for jet-induced star formation.
\item the existence of Minkowski's object and now the companion galaxy HE0450$-$2958 which may be newly formed galaxies induced by QSO activity. 
\item the radio-optical alignment of host galaxies.
\item the common presence of EELRs around radio quasars as candidate seeds for future galaxy formation. 
\item the positional offset of molecular gas concentrations with respect to QSOs. 
\item converging evidence that the M$_{\rm BH}/$M$_{\star}^{\rm bulge}$ ratio was larger in the past, implying that SMBH formed first, then possibly influencing the stellar mass building of host galaxies.
\end{itemize}
We discussed the possibility that EELRs may represent candidate proto-galaxies that will start forming stars after their nearby quasar will have stopped photoionizing them or even before if they are hit by a radio jet. They might therefore not only participate to the building of host galaxies but also galaxies in the nearby environment of massive galaxies. Finally, the last two items of this list (CO offset and large M$_{\rm BH}/$M$_{\star}^{\rm bulge}$ ratio in distant galaxies) are proposed as two independant tests for future observations on the role of radio jets emitted by QSOs (and AGNs) in the formation of new galaxies. In HE0450$-$2958, the bulk of the molecular gas is found associated with the offset companion galaxy and with a radio jet. Several other systems are listed showing similar properties, i.e. an offset between the QSO and peak CO emission, which may be explained by the presence of a radio jet. More statistics on black hole and host galaxy masses in distant QSOs should soon tell us which one formed first or if both followed a parallel path.
 
A natural extension of this work would be to search for separated quasar and star formation activity from mid infrared imaging in other systems. The use of VISIR remains limited to the local Universe but in the near future, high-resolution mid infrared imaging from the Mid Infra-Red Instrument (MIRI) on-board the James Webb Space Telescope (JWST) or from the mid infrared E-ELT Imager and Spectrograph (METIS) project instrument for the ESO-ELT (extremely large telescope) will be extremely powerful to study the connection between galaxy formation and supermassive black hole growth. In the closer future, the generalized study of offset CO emission and of the cold dust distribution will become feasible thanks to the Atacama Large Millimeter Array (ALMA). 


\begin{acknowledgements}
We are grateful to the referee, Dr Alvaro Labiano, for his helpful report and suggestions. We also thank I.Feain and P.Papadopoulos for kindly providing us with the ATCA radio continuum and CO images of the field of HE0450$-$2958 and E. Audit for fruitful discussion. KJ acknowledges support through the Emmy Noether Programme of the German Science Foundation (DFG) with grant number JA 1114/3-1. This research has made use of the NASA/IPAC Extragalactic Database (NED) which is operated by the Jet Propulsion Laboratory, California Institute of Technology, under contract with the National Aeronautics and Space Administration.
\end{acknowledgements}

\end{document}